\documentclass[prc,twocolumn,amsmath,amssymb]{revtex4}

\usepackage{graphicx}
\usepackage{dcolumn}
\usepackage{bm}
\usepackage{amsmath}
\usepackage{float}

\begin{document}

\title{Elastic Electron Scattering from Deformed and Oriented Odd-A Nuclei}
\author{P.~Sarriguren}
\email{p.sarriguren@csic.es}
\affiliation{Instituto de Estructura de la Materia, IEM-CSIC, Serrano 123, E-28006 Madrid, Spain}
\date{\today}

\begin{abstract}
Interference terms between monopole and quadrupole Coulomb form factors that contribute to the cross-section of electron scattering from polarized nuclei are studied within the plane wave Born approximation. By experimentally exploring the effect of this nuclear response, valuable insights into the sign of the quadrupole deformation can be obtained. The nuclear structure is evaluated numerically using a self-consistent deformed Hartree-Fock approximation with Skyrme effective interactions and accounting for pairing correlations. To illustrate the practical applicability, several examples of odd-A nuclei with different spins and deformations are examined. Comparison of unpolarized and polarized cross-sections becomes a tool to discriminate between the oblate and prolate nature of the nuclear shape, providing a deeper understanding of the nuclear deformation.
\end{abstract}

%keywords: elastic electron scattering, axially deformed nuclei, selfconsistent mean field, %charge form factors
\maketitle

\section{Introduction}

Electron scattering from atomic nuclei is a powerful tool for extracting fine details about nuclear properties, such as sizes, shapes, electromagnetic moments, charge and current distributions, form factors, and other inherent characteristics \cite{e-scatt2,e-scatt3,e-scatt4}. This is possible because the process is driven by the interaction between the electron and the electromagnetic fields of the nuclei, which is described by the well-known Quantum Electrodynamics theory. This great advantage has been exploited in the past providing unprecedented insights into the internal structure of the nucleus. Nevertheless, the potential for further learning about the nuclear structure through electron scattering remains far from exhausted.

The feasibility of studying electron scattering from unstable nuclei is becoming a reality. A significant step in this direction has been achieved through the SCRIT project at RIKEN \cite{unstable,scrit_xe}, which has successfully conducted pioneering electron scattering experiments using a radioactive target, specifically on $^{137}$Cs \cite{riken-137cs}. Looking ahead, the forthcoming ELISe project at GSI-FAIR \cite{unstable,anton_nima} is another example of future facilities of collider experiments with electrons and radioactive ion beams. 

Moreover, the study of the effects of the electroweak interaction in the nucleus has provided researchers with a tool to investigate not only the electromagnetic properties of nuclei, but also those due to the weak interaction. Prominent among these pursuits are the parity-violating electron scattering experiments, namely PREX \cite{prex} and CREX \cite{crex}, conducted at the Jefferson Laboratory, investigating lead and calcium nuclei, respectively. These experiments aim to measure the weak charge distribution of the nucleus, providing insights into the nuclear structure, neutron densities, and even the fundamental constituents, the nucleons themselves.

  From conventional scattering experiments employing fixed unpolarized stable targets to more complex methodologies involving polarized electrons and targets, the field continues to advance rapidly. A notable innovation involves the incorporation of polarized internal targets positioned within the experimental setups, that strategically intercept the circulating electrons in storage rings. This enhances control and manipulation of target polarization during experiments. Various combination possibilities of beam and target polarizations for different purposes have been carried out at prominent facilities like Hall-A at Jefferson Laboratory \cite{pol-jlab}, Mainz Microtron (MAMI/MESA) \cite{pol-mainz}, and RIKEN \cite{pol-riken}. The forthcoming generations of research facilities, such as the Electron-Ion Collider at Brookhaven National Laboratory, have even more promising prospects, determined by the unparalleled quality expected in the polarizations of electrons and ion beams \cite{eic}.

The nuclear charge density distribution obviously influences the electron scattering cross-section by means of the so-called charge form factors \cite{e-scatt2,e-scatt3,e-scatt4}. This relationship has been investigated using different theoretical approaches. These encompass the Shell Model \cite{brown,radhi}, along with both relativistic \cite{wang-prc71,roca-prc78,liu-prc96,liang-prc98} and non-relativistic \cite{richter,anton_2005, sarri_prc76,wang-jpg47} self-consistent mean-field models. All of them show robust correlations between these form factors and the underlying density distributions.
  
Obviously, the form factors contain information about nuclear deformation as well. The deformation of a nucleus refers to its departure from a spherical shape due to the arrangement of protons and neutrons within it. Understanding nuclear deformation is crucial to get information about the underlying nuclear forces and the nature of the nuclear many-body system. A large variety of nuclear properties are understood in terms of nuclear deformation. These include among others, rotational energy spectra, transition probabilities, and electromagnetic moments. In addition, the nuclear shape plays an important role in understanding nuclear decays and reactions, with an impact on other domains such as nuclear astrophysics \cite{sarri-stellar1,sarri-stellar2} and particle physics \cite{2bb-def1,2bb-def2,2bb-def3,scaling}. 

Although a comprehensive characterization of the nuclear shape requires numerous degrees of freedom, reducing them to axial symmetry and quadrupole deformation is a good starting point. This approximation is in most cases sufficient to describe reasonably well the properties of a substantial number of the existing nuclei, particularly those at the focus of this study. Within this simplified framework, the shape of a nucleus can be properly described by the quadrupole deformation parameter ($\beta_2$). Its sign provides information about the ratio of axes. If the symmetry axis is larger (smaller) than the perpendicular axes, the sign of $\beta_2$ is positive (negative) and the nuclear shape is prolate (oblate).

Nuclear deformation is theoretically obtained as the shape configuration with minimum energy. However, empirical data on $\beta_2$ and particularly its sign remain somewhat limited \cite{stone}, which poses a real challenge. Electric quadrupole transition probabilities $B(E2)$ are systematically used to extract the value of $\beta_2$ \cite{pritychenko}, but the quadratic dependence on $\beta_2$ prevents obtaining its sign. In fact, the sign of deformation is a very elusive observable that requires the use of sophisticated experimental methods to get insight into the oblate or prolate character of the deformation. Various experimental methods based on $\gamma$-ray spectroscopy and Coulomb excitation \cite{coulex1}, rotational and vibrational spectroscopy \cite{bm,ring}, isotope shifts \cite{marsh}, and beta-decay studies \cite{nacher,sarri-beta}, have been employed to determine the nuclear deformation. In addition, since electron scattering experiments provide information about the charge and magnetic distributions within the nucleus, it is obvious that they can reveal the presence of nuclear deformation by studying the angular distribution of the scattered electrons and the form factors.

In the case of deformed nuclei, the Coulomb form factors can be conveniently written in terms of multipoles, denoted shortly as $C \lambda$ \cite{e-scatt3,moya}. They are sensitive to the various components of the deformed charge density distribution into a multipole expansion. The $C0$ multipole would contain information about the spherical component, while $C2$ would contain information concerning the quadrupole deformation. However, in standard electron scattering experiments with unpolarized electron beams being scattered by unpolarized nuclei, the total charge form factor contributes to the cross-section as an incoherent summation of squared multipoles. 
Hence, the observables sensitive to nuclear deformation are hidden due to two key factors. Firstly, the contributions of the higher-order multipoles to the cross-section decrease fast with increasing multipolarity. Secondly, solely the squared values of $C2$ multipoles contribute, preventing the extraction of information regarding the sign indicative of the oblate or prolate character of the deformation. Consequently, due to these reasons, information about nuclear deformation is difficult to disentangle.
  
A viable strategy to overcome this challenge involves using polarization degrees of freedom in these scattering processes \cite{moya,weigert,donnelly}. In the experiments with polarized beams and/or targets, polarization observables provide valuable information about the nuclear structure, which is hidden in the unpolarized case. In particular, new observables emerge that are sensitive to nuclear deformation, enabling studies about the anisotropy of charge and magnetic distributions within the nucleus. These novel observables contain terms of interference between the different multipoles that can be isolated experimentally by choosing properly the kinematic conditions and the nuclear polarizations.

This work focuses on the new information that can be extracted via scattering of unpolarized electron beams that interact with nuclei intentionally oriented along a quantization axis. Novel and remarkable opportunities arise in this case to gain deeper insights into the underlying nuclear structure. This study was started in a previous work \cite{pol-sarri}, where the suitability of the method to obtain information about the deformation sign was demonstrated through selected examples. Here, a more comprehensive and systematic study is performed, including an array of different nuclear spins ($I=$3/2, 5/2, 7/2, 9/2) and deformation types (oblate, prolate or shape coexistence). This exploration offers a more complete understanding of the interplay between nuclear spins, deformation characteristics, and the observable outcomes of electron scattering experiments, focusing finally on the distinct signatures arising from the different deformation patterns.

The following sections II and III introduce the theoretical framework employed to describe both the reaction mechanism and the nuclear structure involved in electron scattering from deformed and aligned odd-$A$ nuclei. Section IV presents the outcomes of the investigation, focusing on various nuclei with different spin-parities in their ground states that also exemplify different types of deformation. Concluding remarks are shown in section V.

\section{Electron scattering cross-section from aligned targets}

The theoretical framework for describing inclusive electron-nucleus scattering involving polarization degrees of freedom in both  the incident electron beams and the target nuclei, has been introduced in earlier studies \cite{weigert,donnelly}. This framework has further been adopted to address the specific scenario of deformed nuclei as detailed in Refs. \cite{moya,garrido}. The formalism is based on the simplest theoretical framework, the plane wave Born approximation (PWBA), which assumes that a single photon is exchanged in the process and that electrons are properly described as plane waves. 

This approximation provides a straightforward and intuitive way to interpret Coulomb form factor measurements by linking them to the Fourier transforms of the charge matrix elements. Proper treatment of the electron distortion caused by the nuclear Coulomb field requires more involved calculations, such as the distorted wave Born approximation (DWBA) \cite{yennie}. However, the influence of these distortions is relatively minor in the context of this study. This is primarily due to their limited impact at low momentum transfers, typically  below $q\leq 1.2$ fm$^{-1}$, where the effect of the target orientation is already substantial.

Here, we adopt the formalism and notation established in Ref. \cite{moya} for the case of unpolarized electrons scattered by deformed oriented nuclei, which are characterized by a ground state with spin $I_i$ and parity $\pi_i$ ($I_i^{\pi_i}$), leading to a final state $I_f^{\pi_f}$. The cross-section can be expressed as,
\begin{equation}
  \left.  \sigma_{\rm tot} (\theta', \phi')  \right| _{I_i^{\pi_i}\rightarrow I_f^{\pi_f}}
  = Z^2 \sigma_M f_{\rm rec}^{-1} \left[ \sigma_0 + \sigma_{\rm al} (\theta', \phi') \right] \, ,
\end{equation}
in terms of the Mott cross-section $\sigma_M$ and the recoil factor $f_{\rm rec}$. The angles $(\theta', \phi')$ define the polarization of the target relative to the direction of the momentum transfer $\vec{q} =\vec{k_i}-\vec{k_f}$. 

The cross-section $\sigma_0$ is independent on polarizations and can be written in terms of longitudinal ($L$) and transverse ($T$) form factors carrying the information about the charge distribution and the electric and magnetic currents, respectively. It is written as,
\begin{equation}
  \sigma_0 = V_L \left| F_L(I_i,I_f;q) \right| ^2 + V_T \left| F_T(I_i,I_f;q) \right| ^2 \,  .
  \label{sigma-0}
\end{equation}

Similarly, the new cross-section that appears when the target nuclei are aligned is independent of projectile polarizations and given by
\begin{align}
\sigma_{\rm al} (\theta', \phi')   =  \sum_{\ell =even>0} & \alpha_{\ell}^{I_i}  [ P_{\ell} (\cos \theta') \left( -V_L   F_L^{\ell} + V_T F_T^{\ell} \right)  \nonumber \\
& + P_{\ell}^2 (\cos \theta') \cos (2\phi') V_{TT}  F_{TT}^{\ell} \nonumber \\
  & + P_{\ell}^1 (\cos \theta') \cos \phi' V_{TL}  F_{TL}^{\ell} ]  \, .
\label{sigma-al}
\end{align}
The kinematic factors $V_L$ and $V_T$ are given by,
\begin{equation}  
V_L=(Q^2/q^2)^2\ , \quad  V_T=\tan^2 (\theta_e/2) - (Q^2/q^2)/2\, ,
\end{equation}
where $Q^{\mu}=(\omega,\vec{q})$ stands for the four-momentum transferred to the nucleus in the process where an incident electron with four-momentum $k_i ^{\mu}  =(\varepsilon_i,\vec{k_i})$ is scattered through an angle $\theta_e$ to an outgoing electron with  four-momentum $k_f ^{\mu}  =(\varepsilon_f,\vec{k_f})$, with $\omega=\varepsilon_i -\varepsilon_f$ and $\vec{q}=\vec{k_i}-\vec{k_f}$. 

The terms labeled $TT$ and $TL$ contain interferences between transverse-transverse and tranverse-longitudinal multipoles, respectively. They are not further taken into account here because their contribution to the cross-section is canceled when the nuclei are polarized in the $\theta'=0$ direction, which for simplicity will be the only case studied here. The terms $L$ and $T$ can be separated with suitable choices of the kinematic variables using standard Rosenbluth methods.

The structure functions $F_L^{\ell}$'s in Eq. (\ref{sigma-al}) contain terms involving products of different charge multipoles. They can be separated using the dependence of the cross-section on the direction of polarization $(\theta',\phi')$, on the population of the states, as well as on the scattering angle $\theta_e$. 

The statistical tensors $\alpha_{\ell}^{I_i}$ describe the degree of polarization of the nucleus, taking into account the fact that that the $2I_i+1$ sub-states are not equally populated,
\begin{equation}
\alpha_{\ell}^{I_i} = \sum_{M_i} P(M_i) \langle I_i M_i \ell 0 | I_i M_i \rangle \, .
\label{stat}
\end{equation}
$P(M_i)$ are the population probabilities of the target sub-states $|I_iM_i>$ along the direction of polarization  $(\theta', \phi')$. With this definition, $\alpha_{\ell =0}^{I}=1$ for $\ell =0$, irrespective of the state of polarization and $\alpha_{\ell}^{I}=\delta_{\ell,0}$ for unpolarized targets where $P(M_i)=1/(2I_i+1)$. Furthermore, all the statistical tensors with odd $\ell$ values vanish in the case of aligned nuclei ($P(M_i)= P(-M_i)$).
In such scenarios, the use of polarized electrons makes no difference.

Focusing on the longitudinal form factors, also referred as charge or Coulomb form factors, which are the object of this study, their contribution to the unpolarized cross-section can be expressed as a sum of charge multipole form factors $F^{C\lambda}(q)$ as,
\begin{equation}
\left| F_L(I_i,I_f;q) \right| ^2 \, = \sum _ {\lambda \ge 0}  \left| F^{C\lambda}(I_i,I_f;q) \right| ^2 \, ,
\end{equation}
with
\begin{equation}
  F^{C\lambda}(I_i,I_f;q) = \frac{\sqrt{4\pi }}{Z}\langle I_f || \hat T ^{C\lambda}(q)  || I_i \rangle /\sqrt{2I_i+1}\, .
\end{equation}
Hence, the various charge multipole contributions add incoherently, sharing identical factors within the cross-section, thus preventing their kinematic separation. 

The Coulomb multipole operators are given by
\begin{equation}
\hat T ^{C\lambda} _\mu (q) = i ^{\lambda} \int d{\bf R} j_{\lambda}(qR) Y^{\mu} _\lambda (\Omega_R) \hat \rho ({\bf R}) \, ,
\end{equation}
as functions of the nuclear charge operator $\hat \rho ({\bf R})$ to be specified later within the nuclear structure model used.

In contrast to $F_L$, the new form factors $ F_L^{\ell} $ in $\sigma_{\rm al}$ contain terms of interference between Coulomb multipoles:
\begin{equation}
F_L^{\ell} (I_i,I_f;q) = \sum _{\lambda, \lambda'} X(\lambda, \lambda', I_i, I_f, \ell )
F ^ {C\lambda}(q) F ^ {C\lambda'}(q)\, ,
%F ^ {C\lambda}(I_i,I_f;q) F ^ {C\lambda'}(I_i,I_f;q)\, ,
\label{interfff}
\end{equation}
where $X(\lambda, \lambda', I_i, I_f, \ell )$ are factors that involve 3j- and 6j- Wigner coefficients \cite{moya},
\begin{equation}
X = (-1)^{I_i+I_f+1} {\bar I_i}{\bar \lambda} {\bar \lambda '}{\bar \ell}^2 
\left(  \begin{array}{ccc}  
                 \lambda   &  \lambda '   &  \ell  \\   0 &  0 &  0
           \end{array}   \right) 
\left\{  \begin{array}{ccc}
                 I_i   &  I_i   &  \ell  \\
                \lambda &  \lambda ' &  I_f
             \end{array}     \right\} \, ,
\label{eq-x}
\end{equation}
with $\bar a=\sqrt{2a+1}$.

The form factors $F_L^{\ell}$'s  can be isolated by measuring the cross-sections at $\theta' =0$ that corresponds to nuclei oriented along the  $\vec{q} =\vec{k_i}-\vec{k_f}$ direction,
\begin{equation}
  \sigma_{\rm al} = \sum_{\ell} \alpha_{\ell}^{I_i}  \left( -V_L   F_L^{\ell} + V_T F_T^{\ell} \right) = \frac{\sigma_{\rm tot} (\theta'=0,\phi')}{
  Z^2 \sigma_M f_{rec}^{-1}} -\sigma_0 \, .
\end{equation}
Using the kinematic dependence of the $V_L$ and $V_T$ factors, $F_L^{\ell}$ and $F_T^{\ell}$
can be isolated and studied individually.

In rotational nuclei, the multipole charge form factors corresponding to a given transition $I_ik \rightarrow I_fk$ within a band characterized by the projection $k$ of the total spin along the symmetry axis, can be expressed in terms of the so-called intrinsic form factors ${\cal F} ^{C\lambda}$. The expression that relates laboratory and intrinsic charge multipole form factors, to lowest order in angular momentum \cite{moya}, is given by
\begin{equation}
F^{C\lambda}(I_i,I_f;q) = \langle I_i k \lambda 0 | I_f k \rangle {\cal F} ^{C\lambda}(q) \, ,
\label{intrinsic}
\end{equation}

\begin{equation}
{\cal F} ^{C\lambda} (q) = i^{\lambda} \sqrt{\frac{4\pi}{2\lambda+1}}  \int_0^\infty R^2dR \rho_\lambda(R) j_\lambda (qR)\, .
\label{f-intrin}
\end{equation}
The multipole components of the density distribution $\rho_\lambda(R)$ will be specified later as the coefficients of its expansion into Legendre polynomials.
Center of mass corrections are included in the harmonic-oscillator approximation with a factor $\exp [q^2/(4A^{2/3})]$. 
Moreover, finite size effects are included as a sum of monopoles \cite{simon} for protons and by the difference of two Gaussians \cite{chandra} for neutrons.

The main objective of this work is the study of the interference between the monopole and quadrupole charge form factors, as they appear in Eq. (\ref{interfff}). The ground states of even--even rotational nuclei, characterized by $(I=k=0)$, lack the capacity of being polarized and therefore, $C2$ multipoles can only be extracted in a quadratic form as $|F^{C2}|^2$. This prevents determination of the deformation sign. This extraction can be achieved from inelastic transitions to the $2^+$ excited states. In contrast, apart from nuclei with spins $I=1/2$, odd-$A$ rotors can be aligned, leading to the emergence of interference contributions to the cross-section. Therefore,  only elastic electron scattering from nuclei with spins $I \ge 3/2$ will receive contributions from both $C0$ and $C2$ and their interference can be studied. Furthermore, only $\ell =2$ terms are needed to study such interference phenomena.

In the case of elastic scattering $I_i=I_f=k$, the notation for charge multipoles simplifies to $F^{C\lambda} (I_i=I_f,I_f;q) \equiv F^{C\lambda}_{I_f}(q)$, and the contribution from aligned nuclei reduces to

\begin{equation}
\sigma _{\rm al} = - \alpha_2  ^{I_i} P_2 (\cos \theta') V_L F_L  ^{\ell=2}\, .
\label{k32}
\end{equation}
In this case and for $\ell=2$, the $X$ coefficient in Eqs. (\ref{interfff}) and (\ref{eq-x}) is given by $X=-2\sqrt{5}$, independent of the spin values. Then,
\begin{equation}
F_L^{\ell=2}(I_f;q) = -2\sqrt{5} F_{I_f}^{C0}(q) F_{I_f}^{C2}(q) \, .
\end{equation}

To evaluate the statistical tensors, we can assume the simplest scenario where the target is oriented along the momentum transfer direction ($\theta'=0$) and the alignment is realized as $P(M_i=+I_i)=P(M_i=-I_i)=0.5$, or equivalently for fully polarized nuclei $P(M_i)=\delta_{M_i,+I_i}$. In this case the $\alpha_2 ^{I_i}$ factors in Eq. (\ref{stat}) take specific values, namely $ 1/ \sqrt{5}$, $\sqrt{5/14}$, $\sqrt{7/15}$, and $\sqrt{6/11}$, for $I_i=3/2$, $5/2$,  $7/2$, and $9/2$, respectively.

Under these conditions, apart from  kinematic factors, the contribution of the longitudinal terms  to the total cross-section can be written as 
\begin{equation}
 \sigma_{\rm tot} \sim \left( \sigma_0 + \sigma_{\rm al} \right)  \sim V_L F_{\rm eff}\, ,
\end{equation}
where
\begin{equation}
  F_{\rm eff} =  |F_L|^2 + F_{02} =  |F^{C0}|^2 + |F^{C2}|^2 + F_{02} \, ,
  \label{feff}
\end{equation}
with
\begin{equation}
  F_{02} = {\cal A} (I_f){\cal F}^{C0}(q) {\cal F}^{C2}(q) \, ,
\label{f02}
\end{equation}
with ${\cal A} = 2/\sqrt{5}$, $5\sqrt{5}/7$, $14/(3\sqrt{5})$, $12\sqrt{5}/11$, for $I_i=I_f=3/2,\, 5/2,\, 7/2,$ and $9/2$, respectively.

\section{Nuclear Structure}

The nuclear structure framework needed to calculate the density distributions involved in the expressions of the multipole charge form factors is derived from a self-consistent deformed Hartree-Fock (HF) calculation with effective Skyrme interactions and pairing correlations in the BCS approximation, as described in Ref.  \cite{vautherin}. Specifically, the Skyrme SLy4 interaction \cite{skyrme}, along with phenomenological pairing gaps, are used in these calculations. This formalism has previously found application in electron scattering studies from axially deformed nuclei, including the computation of both longitudinal and transverse form factors  \cite{kowalski,sarri_ff,nk,ff1,ff2}.
Calculations with other Skyrme interactions have also been performed in those references leading to similar results.

In this approach the nuclear density can be written as,
\begin{equation}
\rho ({\bf R})= 2\sum_i v_i^2 \left| \Phi_i({\bf R})\right| ^2 \, ,
\end{equation}
where $v_i^2$ denote the occupation probabilities and $\Phi_i$ the single-particle Hartree-Fock wave functions. The latter are expanded into the eigenstates of an axially deformed harmonic oscillator potential, using a set of 11 major shells. The harmonic oscillator parameters are properly chosen to minimize the energy in such basis. The quadrupole deformation parameter $\beta_2$ is obtained self-consistently from  the intrinsic quadrupole moment $Q_0$ and the mean square radius $<R ^2>$, both calculated microscopically as functions of the density,
\begin{eqnarray}
  \beta_2 &=& \sqrt{\frac{\pi}{5}}\frac{Q_0}{A<R ^2>}\, , \label{beta2} \\
  Q_0 &=& \sqrt{16\pi/5}\int \rho({\bf R})R^2 Y_{20}(\Omega_R) d{\bf R}\, ,  \\
  \langle R^2 \rangle &=&  \frac{ \int R^2 \rho({\bf R}) d{\bf R}} {\int \rho({\bf R}) d{\bf R}} \, .
\end{eqnarray}

The axially deformed density distribution  $\rho ({\bf R})$ is conveniently expressed through an expansion in Legendre polynomials \cite{vautherin},

\begin{equation}
\rho ({\bf R})= \rho(R\cos \theta, R\sin \theta )=\sum_\lambda \rho_\lambda(R) 
P_\lambda(\cos \theta) \, ,
\end{equation}
where the multipole coefficients are given by,

\begin{equation}
  \rho_{\lambda}(R)= (2\lambda +1) \int _{0}^{+1}  P_{\lambda}(\cos \theta)
  \rho (R \cos \theta , R \sin \theta) d(\cos \theta) \, .
\end{equation}

\begin{figure}[H]
\centering
\includegraphics[width=65mm]{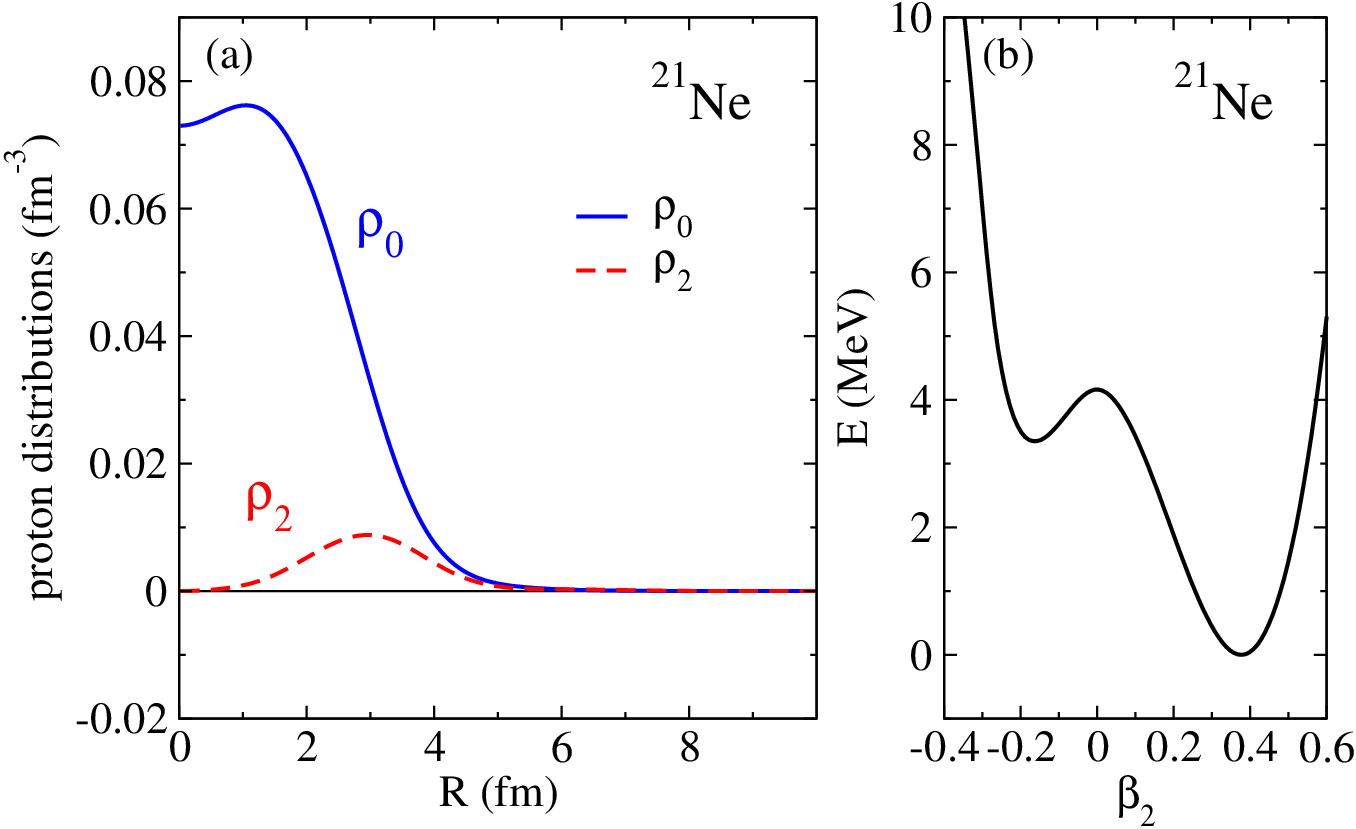}
\caption{(a) Multipole components $\lambda=0,2$ of the proton distributions in $^{21}$Ne $(I^{\pi}=k^{\pi}=3/2^+)$ for the prolate shape configuration. (b) Excitation energy as a function of the quadrupole deformation parameter $\beta_2$.}
\label{fig1_21ne}
\end{figure}

\begin{figure}[H]
  \centering
  \includegraphics[width=65mm]{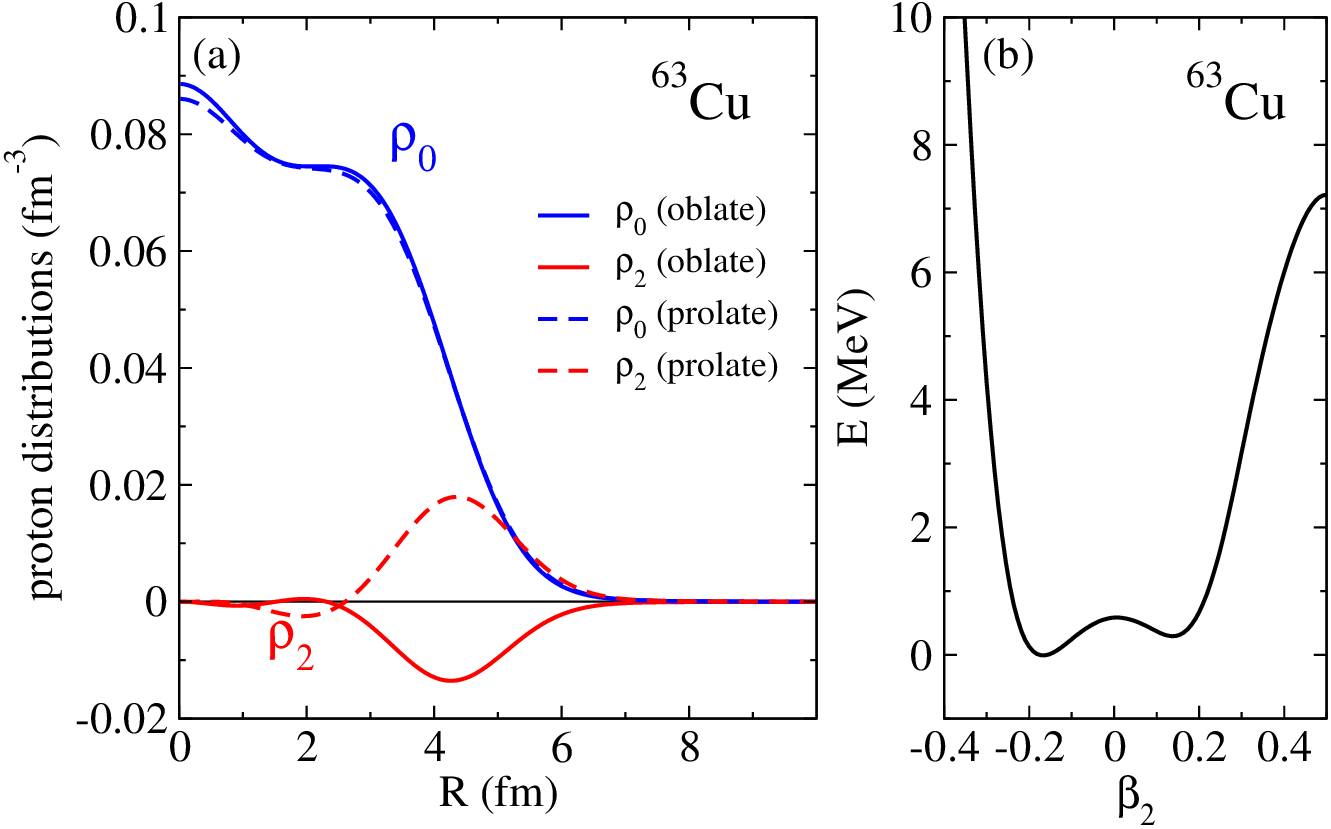}
\caption{Same as in Fig. \ref{fig1_21ne}, but for the oblate and prolate shapes in $^{63}$Cu $(I^{\pi}=k^{\pi}=3/2^-)$.}
\label{fig1_63cu}
\end{figure}

\begin{figure}[H]
  \centering
  \includegraphics[width=65mm]{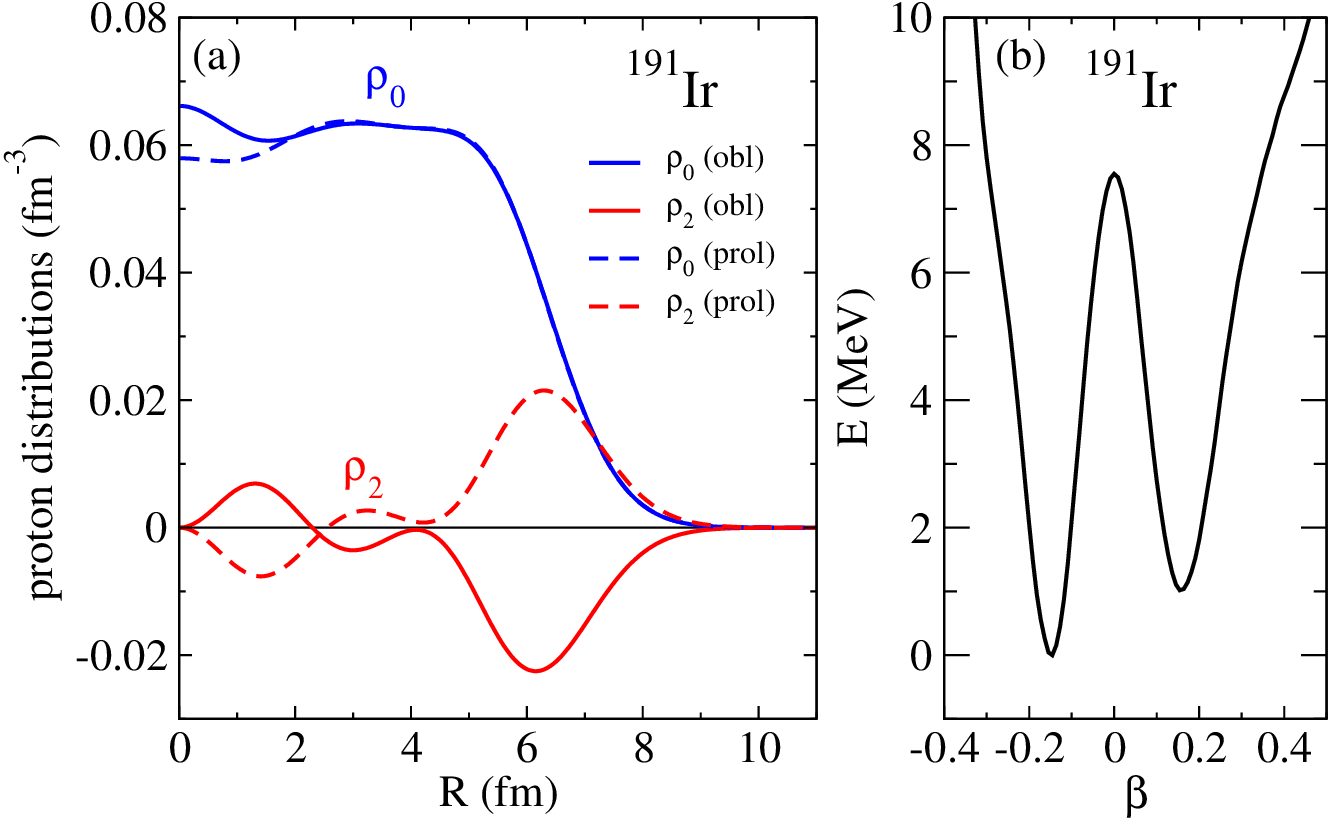}
\caption{Same as in Fig. \ref{fig1_21ne}, but for the oblate and prolate shapes in $^{191}$Ir $(I^{\pi}=k^{\pi}=3/2^+)$.}
\label{fig1_191ir}
\end{figure}

Figures \ref{fig1_21ne}, \ref{fig1_63cu}, and \ref{fig1_191ir} depict the results for the multipole densities $\rho_0$ and $\rho_2$ of the proton distributions in $^{21}$Ne, $^{63}$Cu, and $^{191}$Ir, respectively. Additionally, the figures provide plots of the energy-deformation curves, i.e., binding energies as a function of the quadrupole deformation parameter $\beta_2$ given in Eq. (\ref{beta2}). The chosen nuclei are instances of $I^{\pi}=3/2^+$, $3/2^-$, and $3/2^+$, respectively.

In the case of $^{21}$Ne, only the prolate deformation is considered for further calculations of form factors because its energy lies much deeper than the energy of the oblate configuration. On the other hand, in the other two cases for $^{63}$Cu and $^{191}$Ir, both oblate and prolate shapes are studied because the minima are located at close energies, although in two different scenarios. The relatively shallow minima in $^{63}$Cu potentially indicate an example of soft nucleus, while the pronounced energy minima separated by high energy barriers in $^{191}$Ir, point towards a paradigmatic example of shape coexistence.

In the case of $^{63}$Cu ($Z=29$), the spin-parity assignment of the ground state according to the Nilsson-like diagram in the oblate sector agrees with the experimentally observed $3/2^-$ state. However, the prolate deformation is minimized with the odd proton occupying a $1/2^-$ state originated in the $p_{3/2}$ spherical shell. It is worth noting that a $1/2^-$ state is observed experimentally at an energy $E=693$ keV. In this study, the prolate configuration is constrained to be in a $3/2^-$ state to compare both oblate and prolate configurations with the same spin-parity assignments. On the other hand, the oblate and prolate configurations in $^{191}$Ir are properly described with $3/2^+$ states in accordance with experiment.

Inspection of these figures reveals that $\rho_{2}(R)$ peaks always in the surface region with positive (negative) values depending on the prolate (oblate) character. This difference has an effect in the $C2$ multipole form factor in Eq. (\ref{f-intrin}), becoming a distinctive marker of the oblate or prolate character of the nuclear shape.

\begin{figure}[H]
  \centering
  \includegraphics[width=65mm]{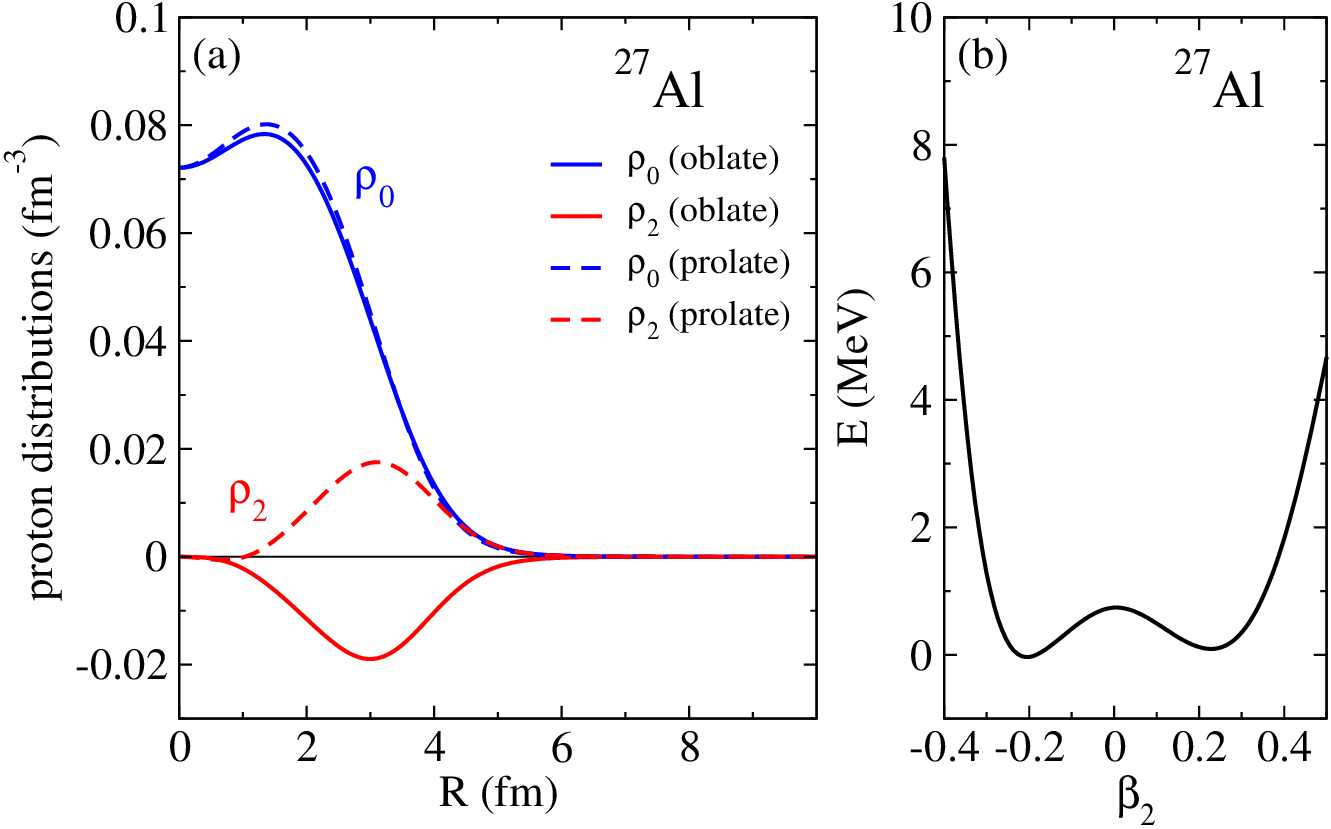}
\caption{Same as in Fig. \ref{fig1_21ne}, but for the oblate and prolate shapes in $^{27}$Al $(I^{\pi}=k^{\pi}=5/2^+)$.}
\label{fig1_27al}
\end{figure}

\begin{figure}[H]
  \centering
  \includegraphics[width=65mm]{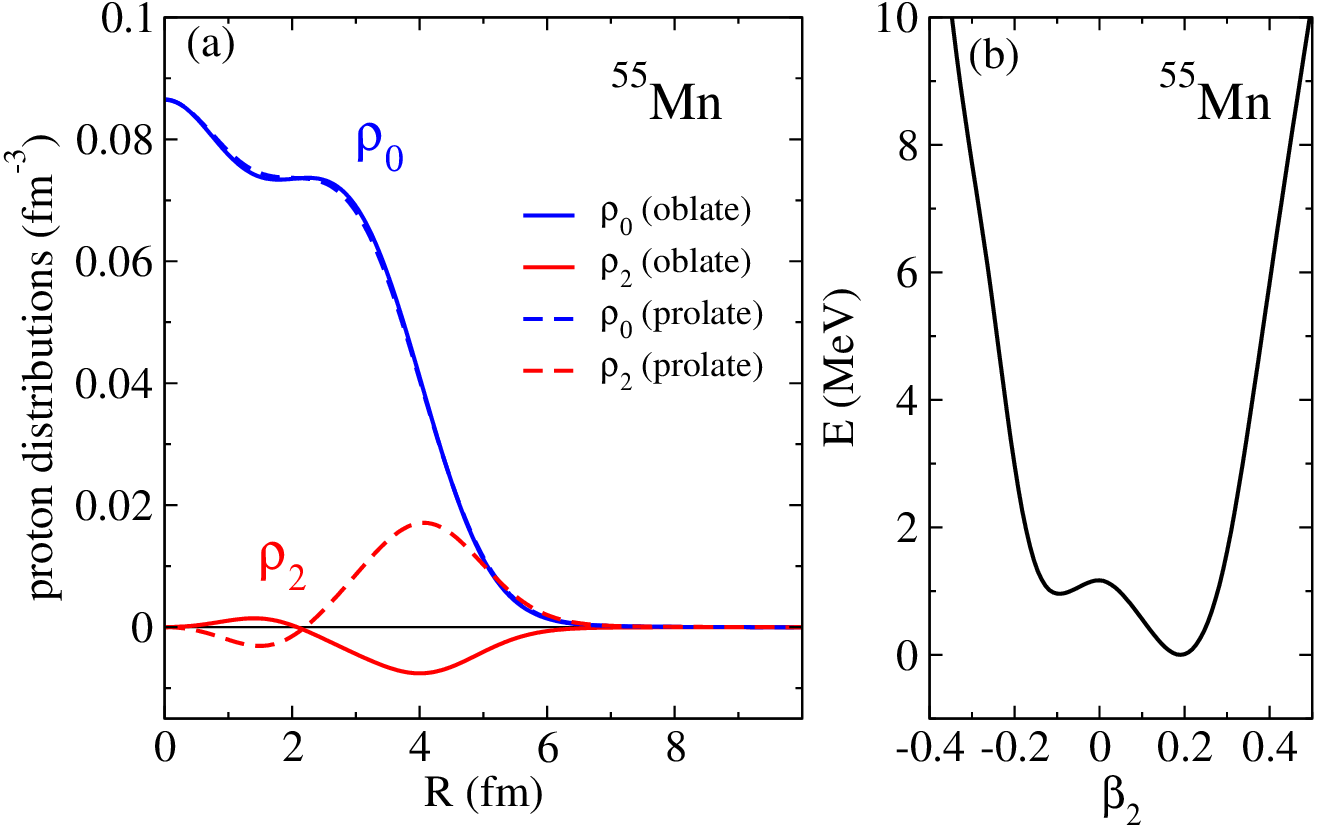}
\caption{Same as in Fig. \ref{fig1_21ne}, but for the oblate and prolate shapes in $^{55}$Mn $(I^{\pi}=k^{\pi}=5/2^-)$.}
\label{fig1_55mn}
\end{figure}

\begin{figure}[H]
  \centering
  \includegraphics[width=65mm]{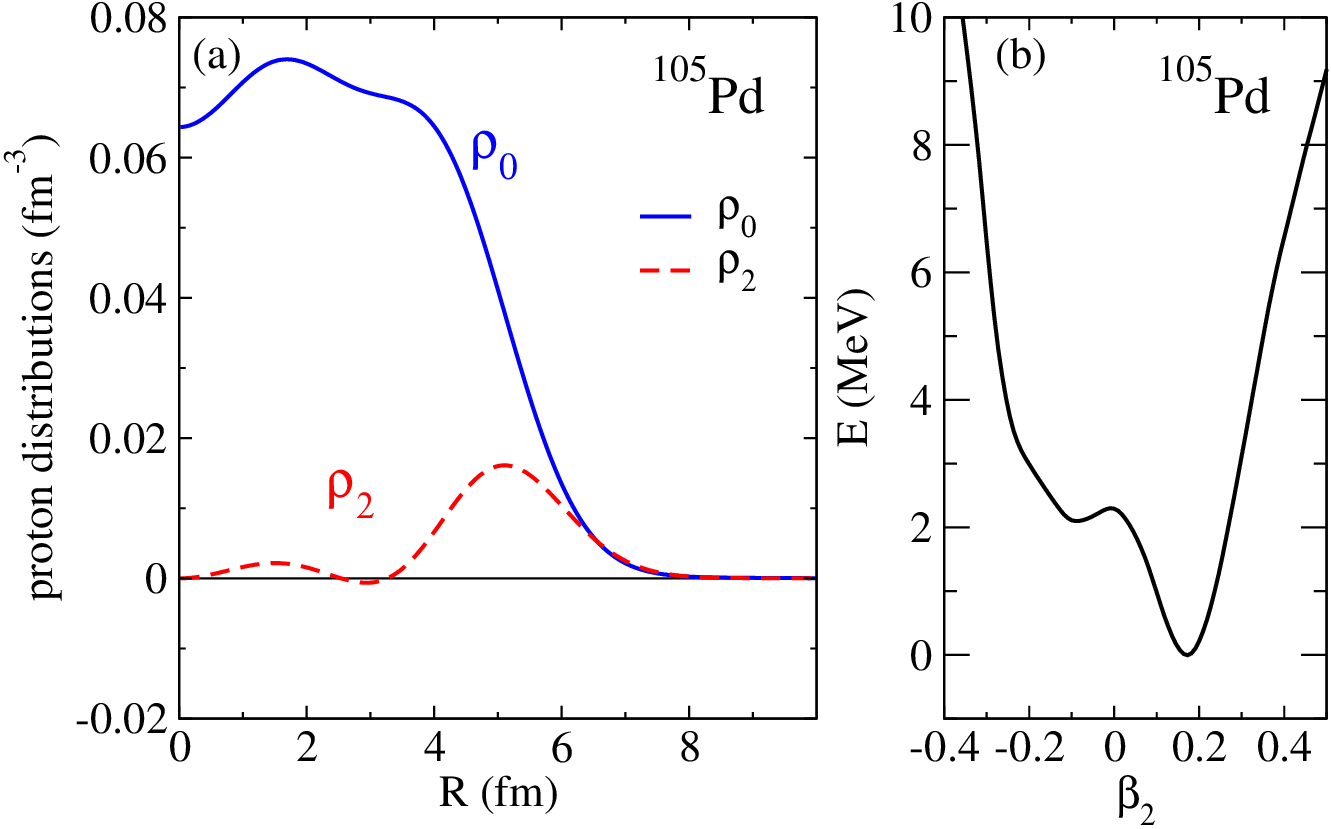}
\caption{Same as in Fig. \ref{fig1_21ne}, but for $^{105}$Pd $(I^{\pi}=k^{\pi}=5/2^+)$.}
\label{fig1_105pd}
\end{figure}

The subsequent set of figures, namely Figs. \ref{fig1_27al}, \ref{fig1_55mn}, and \ref{fig1_105pd}, depicts analogous plots for the cases of $^{27}$Al, $^{55}$Mn, and $^{105}$Pd, which typify examples of $I^{\pi}=5/2^+$, $5/2^-$, and $5/2^+$ nuclei, respectively. For the same arguments as above, oblate and prolate deformations are studied in the cases of $^{27}$Al and $^{55}$Mn, whereas only the prolate case is studied in $^{105}$Pd.
In the case of $^{27}$Al ($Z=13$), the experimentally observed spin and parity of the ground state ($5/2^+$) corresponds to a prolate configuration built on the 
$d_{5/2}$ spherical shell. The oblate configuration is minimized with a $1/2^+$ assignment, which is observed experimentally as an excited state at $E=843$ keV. Similarly to the case of $^{63}$Cu, the oblate solution is constrained to the experimental $5/2^+$. In the case of $^{55}$Mn ($Z=25$), the $5/2^-$ oblate and prolate states from the $f_{7/2}$ shell are used in the calculations.
It is worth noting that for $^{55}$Mn the peaks of $\rho_{2}(R)$ at the surface show the largest difference between oblate and prolate shapes. This feature is related to the larger difference of the magnitude of  $\beta_2$ between the oblate and prolate minima.

The following figures \ref{fig1_167er}, \ref{fig1_177hf}, and \ref{fig1_181ta} correspond to $^{167}$Er, $^{177}$Hf, and $^{181}$Ta, which are examples of $I^{\pi}=7/2^+$, $7/2^-$, and $7/2^+$ nuclei, respectively. Concluding this series, Fig. \ref{fig1_179hf} shows the results for $^{179}$Hf with $I^{\pi}=9/2^+$. In these nuclei the analysis is restricted to the form factors associated with prolate deformations.
The patterns observed in these examples of heavy deformed nuclei with $I=7/2$ and $I=9/2$ show a similar behavior of the density multipoles  $\rho_{0}$ and $\rho_{2}$. The most striking feature regarding the densities is the appearance of wavings in the  $\rho_{2}$ profiles, with large positive peaks on the surface, as expected from prolate configurations. The energy-deformation curves show also a common behavior in those cases with deep minima in the prolate sectors and oblate configurations at excitation energies above 5 MeV that will not be considered further. The sudden change in the spin and parity of the two isotopes $^{177}$Hf ($N=105$) and $^{179}$Hf ($N=107$) from $7/2^-$ to $9/2^+$ should also be highlighted. This is an interesting nuclear structure effect, where the proximity of energy levels in the vicinity of $\beta_2=0.3$ favors a jump between the orbitals of $h_{9/2}$ ($7/2^-$) and $i_{13/2}$ ($9/2^+$).

\section{Results on Form Factors}

The results of this work for the charge form factors $C0$ and $C2$ squared, as well as the interference term $F_{02}$ in Eq. (\ref{f02}) are presented  in Figs. \ref{fig2_21ne} - \ref{fig2_179hf} for the same stable deformed nuclei studied earlier. Namely these figures  contain the results for $^{21}$Ne, $^{63}$Cu, $^{191}$Ir, $^{27}$Al, $^{55}$Mn, $^{105}$Pd, $^{167}$Er, $^{177}$Hf, $^{181}$Ta, and $^{179}$Hf, respectively.

As a general comment, these figures demonstrate the challenge of extracting meaningful information about $C2$ multipoles from unpolarized experiments due to their small magnitude and to their quadratic dependence. In contrast, the interference term $F_{02}$ is significantly greater than $(F^{C2})^2$ and carries information about the sign of deformation. Typically, the first peak of $F_{02}$ is roughly  between one and two orders of magnitude smaller than  $(F^{C0})^2$ and two orders of magnitude larger than  $(F^{C2})^2$.
The first and most prominent peak of the interference term  $F_{02}$ in prolate (oblate) nuclei appears with a negative (positive) sign. Hence, according to its effect on the total cross-section, the latter will be decreased (increased) with respect to the unpolarized cross-section. The strength of this reduction (enhancement) depends on $F_{02}$, which is determined to some  extent by the magnitude of the quadrupole deformation parameter $\beta_2$.  
The study of this dependence makes more sense in cases in which we consider two different deformations in the same nucleus. Specifically, it is observed that in the case of  $^{55}$Mn in Fig.  \ref{fig2_55mn} the differences between the peaks of oblate and prolate form factors in both  $(F^{C2})^2$ and  $F_{02}$ are the largest within the cases studied since the differences in  $\beta_2$ between oblate and prolate deformations in this nucleus are also the largest.

Lastly, turning to Figs. \ref{fig3_21ne} -- \ref{fig3_179hf},
they contain the total longitudinal form factor $(F_L)^2=(F^{C0})^2+(F^{C2})^2$ in the unpolarized case, along with the previously defined $F_{\rm eff}$ form factors for the same illustrative cases discussed earlier.
A comparative analysis between the unpolarized response, which is proportional to $(F_L)^2$, and the polarized counterpart, which is proportional to $F_{\rm eff}$, shows a clear emerging trend. 
By comparing experimental measurements of cross-sections with unpolarized and polarized nuclei, crucial information about the oblate or prolate character of the nuclear shape can be inferred, simply by checking whether the cross-section increases or decreases.
Therefore, the comparison between these two cross-sections provides a signature of the sign of the deformation that is independent of the model.

The magnitude of the deviation provides information about the magnitude of the quadrupole deformation, although in this case in a model-dependent way.
The insets accompanying these figures show the ratios between the cross-sections with aligned nuclei and their unpolarized counterparts,
\begin{equation}
\frac{\left( \sigma_0 + \sigma_{\rm al} \right)}{\sigma_0} = 
1+ \frac{F_{02}}{|F_L|^2} = \frac{F_{\rm eff}}{|F_L|^2} \, .
\end{equation}
These figures show the extent of the deviation from the standard unpolarized cross-section. The divergence is already significant at low momentum transfers and gains prominence when approaching the first diffraction minimum, where the cross-section is still large enough to be measured. Nevertheless, it is important to note that in the close vicinity of the minimum, the cross-section is very small, potentially leading to the emergence of additional effects not included in this study.

\section{Conclusions}

The changes induced by nuclear deformation in the cross-section of electrons scattered by polarized odd-$A$ nuclei have been investigated. Specifically, the study focuses on the effect produced by the interference form factors between monopole and quadrupole Coulomb terms. To illustrate the potential magnitude of those effects, specific instances with different ground-state spins and parities and with different types of deformation are presented. Namely $^{21}$Ne, $^{63}$Cu, $^{191}$Ir, $^{27}$Al, $^{55}$Mn, $^{105}$Pd, $^{167}$Er, $^{177}$Hf, $^{181}$Ta, and $^{179}$Hf. The nuclear structure is described within a deformed self-consistent Skyrme HF+BCS calculations, while the reaction mechanism is based on the PWBA.

The important point is that joint measurements of both the unpolarized and the polarized cross-sections allow one to obtain a model-independent signature of the oblate or prolate deformation character of the nucleus. 
This information is based on the simple observation of whether the cross-section is increased or decreased by polarizing the target nucleus. The magnitude of this deviation is related with the magnitude of the  quadrupole deformation in a model-dependent way.
Although the amplitude of this effect could depend on the specific nuclear model considered, the constraints made in this study do not change the conclusions concerning the feasibility of getting observables that exhibit sensitivity to the $C0/C2$ interference and therefore, to the sign of the axial deformation in odd-A nuclei. 

The nuclear spin in the electron scattering processes studied here determines the number of multipoles contributing to the total form factor. While the primary focus of this paper centers on the $C0/C2$ interference, a broader investigation considering higher multipolarities allowed by the nuclear spin, may reveal additional interferences. Furthermore, the nuclear spin also determines to a large extent the magnitudes of the geometrical factors, dictating the weights of the interference contributions to the total form factor. Conversely, the parity of the nuclear state has no significant influence in this context.
The study encompasses nuclei with different atomic and mass numbers. These numbers play an important role in determining the nuclear size and, specifically, the radius of the nuclear charge. The latter exhibits a direct correlation with the first minimum of the cross section, which undergoes compression with increasing radius. Interestingly, this investigation reveals that for the heavier nuclei under consideration, the quadrupole parameters $\beta_2$ and the density profiles exhibit remarkable similarity, a consequence of the large core involved. This also results in similar outcomes for the $C0/C2$ interference. In contrast, lighter nuclei manifest a larger diversity of shapes, leading to a broader range of density profiles and form factors that depend more on the specific characteristics of each individual example considered.

Future research should address other important aspects, including the investigation of inelastic reactions, the opportunities of polarized electrons, and the impact of the electron Coulomb distortion on the cross-section.

\begin{acknowledgments}
This work was supported by Grant PID2022-136992NB-I00, funded by MCIN/AEI/
10.13039/501100011033 and by “ERDF A way of making Europe”.

\end{acknowledgments}

\newpage

%%%%%%%%%Fig1-1%%%%%%%%%%%%%%%%%%%%%%%%%%%%%%
%%%%%%%%%%%%%%%%%%%%%%%%%%%%%%%%%%%%%%%%%%%%%%%
%%%%%%%%%%Fig1-2%%%%%%%%%%%%%%%%%%%%%%%%%%%%%%
%%%%%%%%%%%%%%%%%%%%%%%%%%%%%%%%%%%%%%%%%%%%%%%
%%%%%%%%Fig1-3%%%%%%%%%%%%%%%%%%%%%%%%%%%%%%
%%%%%%%%%%%%%%%%%%%%%%%%%%%%%%%%%%%%%%%%%%%%
%%%%%%%%%Fig1-4%%%%%%%%%%%%%%%%%%%%%%%%%%%%%%%%
%%%%%%%%%%%%%%%%%%%%%%%%%%%%%%%%%%%%%%%%%%%%
%%%%%%Fig1-5%%%%%%%%%%%%%%%%%%%%%%%%%%%%%%%%
%%%%%%%%%%%%%%%%%%%%%%%%%%%%%%%%%%%%%%%%%%%%
%%%%%%%%%%Fig1-6%%%%%%%%%%%%%%%%%%%%%%%%%%%%%%%%
%%%%%%%%%%%%%%%%%%%%%%%%%%%%%%%%%%%%%%%%%%%%%%%%%%%
%%%%%Fig1-7%%%%%%%%%%%%%%%%%%%%%%%%%%%%%%%%
\begin{figure}[H]
  \centering
  \includegraphics[width=65mm]{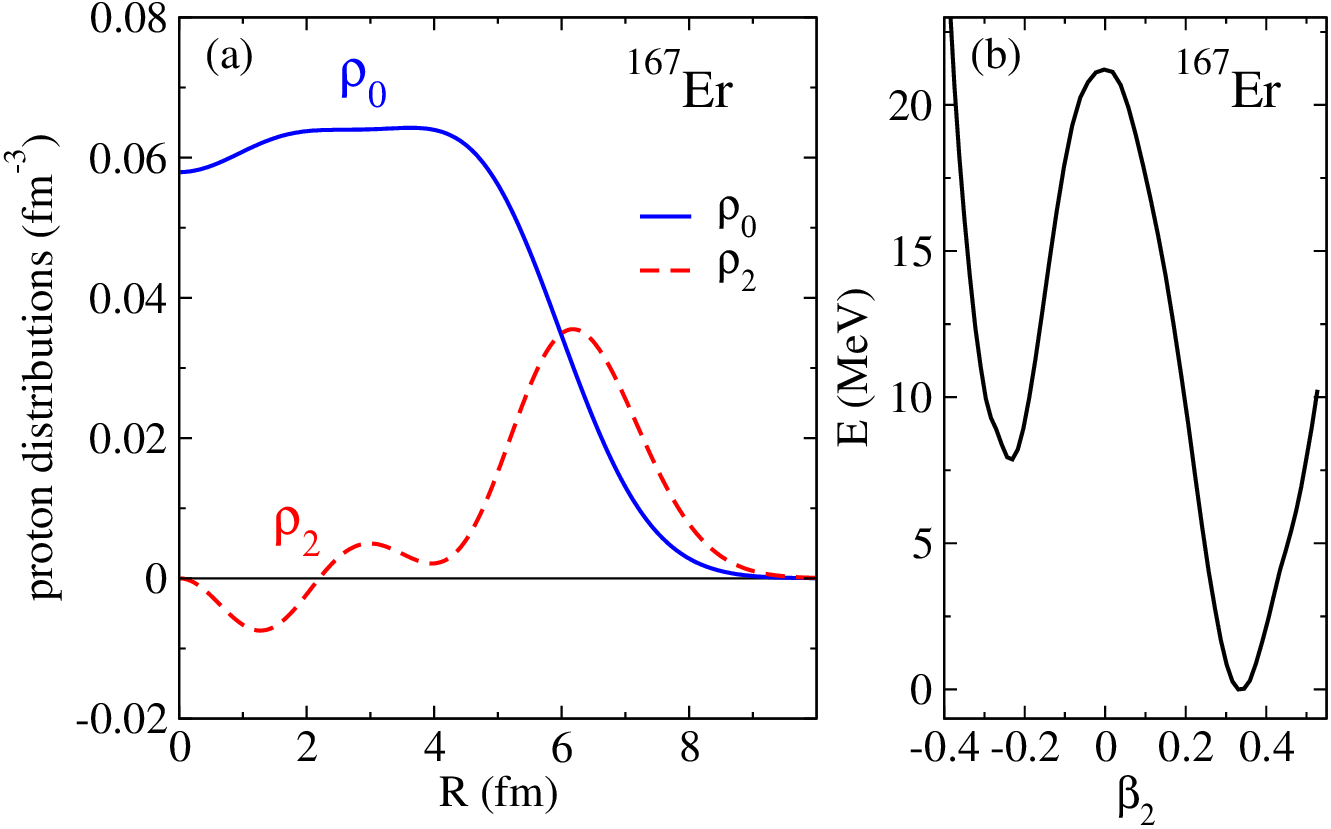}
\caption{Same as in Fig. \ref{fig1_21ne}, but for $^{167}$Er $(I^{\pi}=k^{\pi}=7/2^+)$.}
\label{fig1_167er}
\end{figure}
%%%%%%%%%%%%%%%%%%%%%%%%%%%%%%%%%%%%%%%%%
%%%%%Fig1-8%%%%%%%%%%%%%%%%%%%%%%%%%%%%%%%
\begin{figure}[H]
  \centering
  \includegraphics[width=65mm]{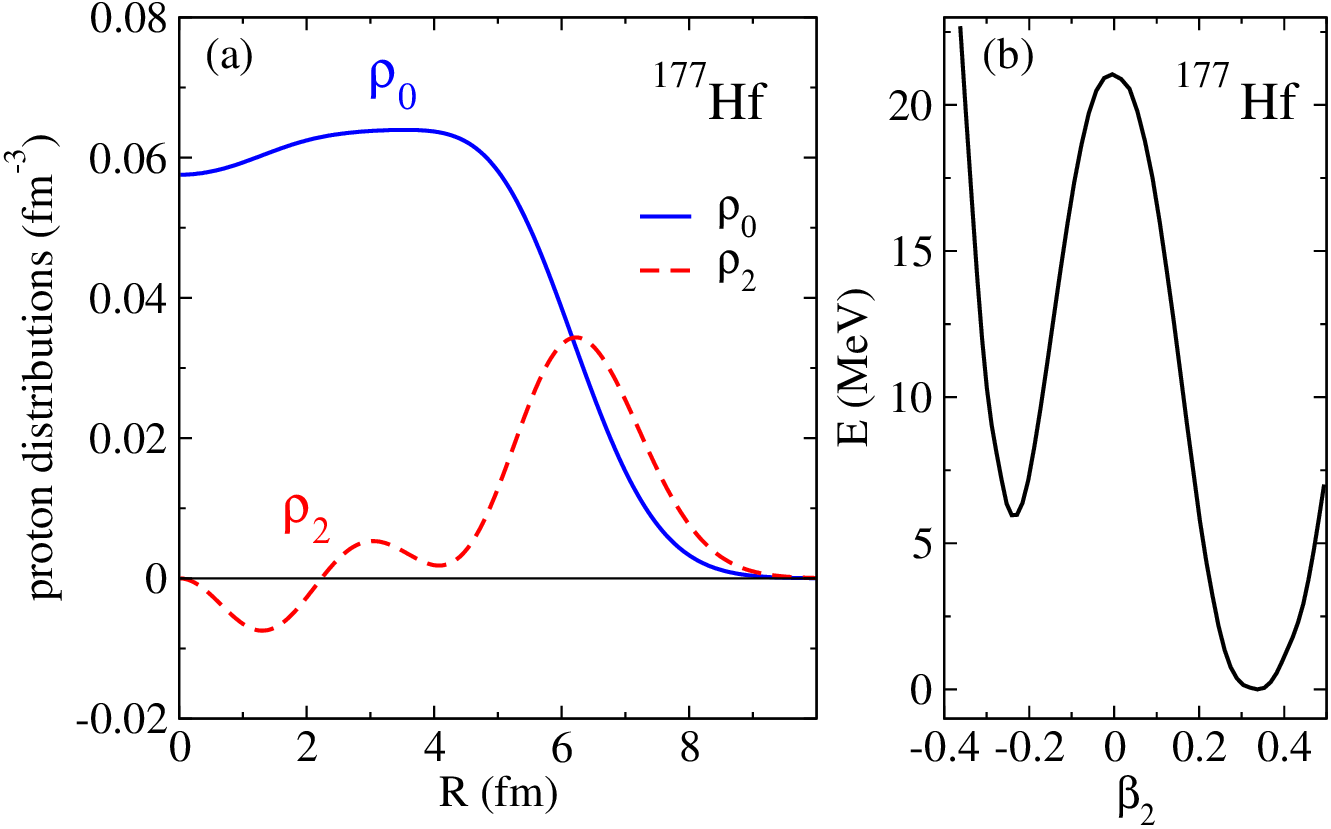}
\caption{Same as in Fig. \ref{fig1_21ne}, but for $^{177}$Hf $(I^{\pi}=k^{\pi}=7/2^-)$.}
\label{fig1_177hf}
\end{figure}
%%%%%%%%%%%%%%%%%%%%%%%%%%%%%%%%%%%%%%%%
%%%%%%%%%%Fig1-9%%%%%%%%%%%%%%%%%%%%%%%%%%%%%%%%
\begin{figure}[H]
  \centering
  \includegraphics[width=65mm]{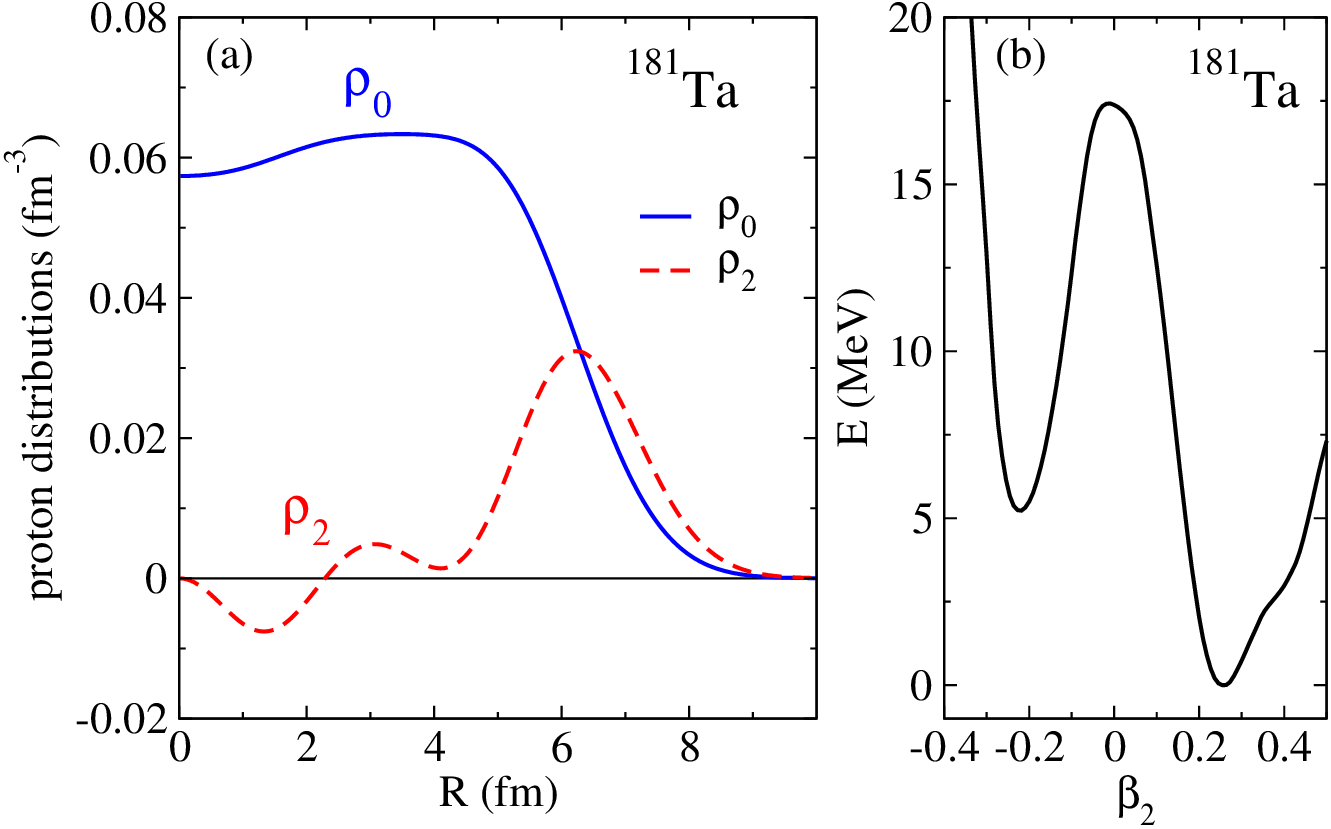}
\caption{Same as in Fig. \ref{fig1_21ne}, but for $^{181}$Ta $(I^{\pi}=k^{\pi}=7/2^+)$.}
\label{fig1_181ta}
\end{figure}
%%%%%%%%%%%%%%%%%%%%%%%%%%%%%%%%%%%%%%%%%%%%%
%%%%%%%%Fig1-10%%%%%%%%%%%%%%%%%%%%%%%%%%%%%%
\begin{figure}[H]
  \centering
  \includegraphics[width=65mm]{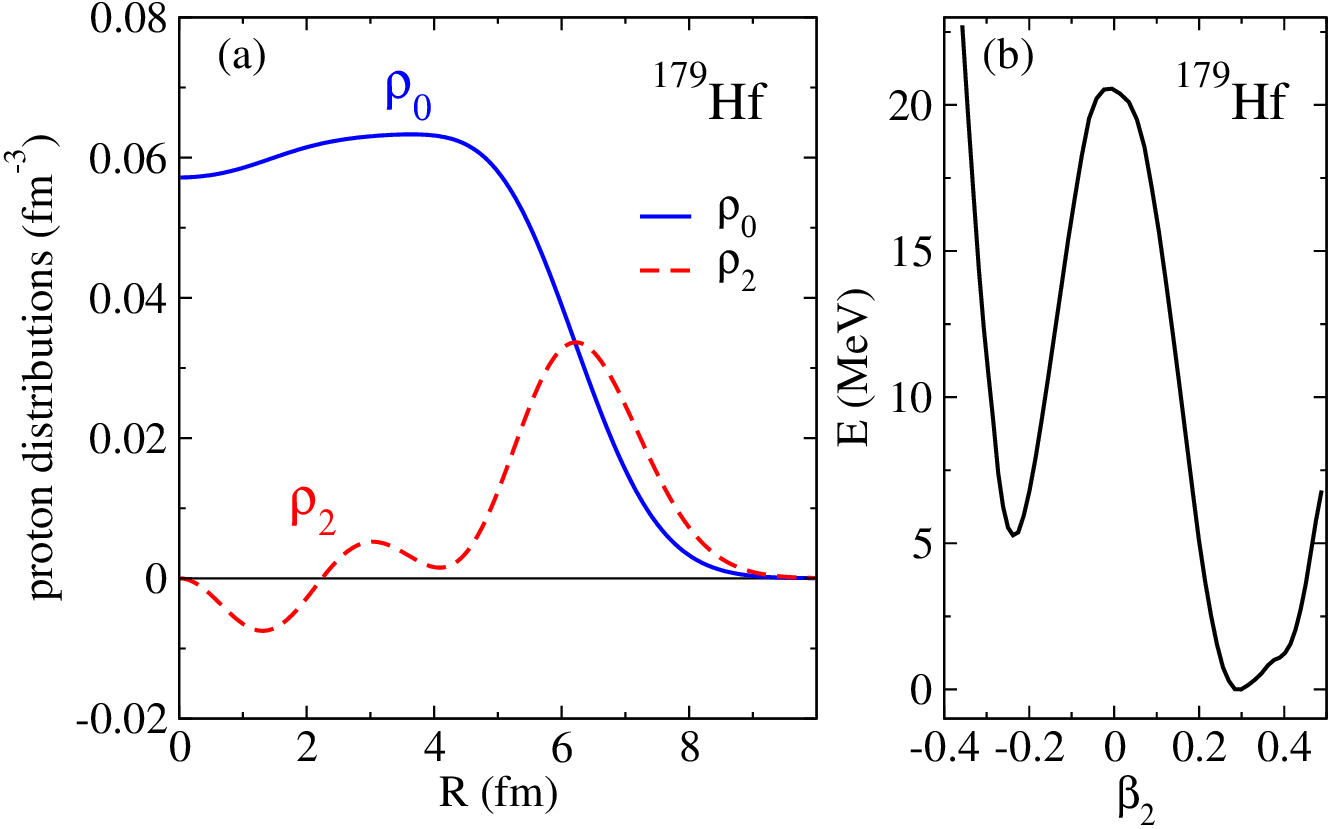}
\caption{Same as in Fig. \ref{fig1_21ne}, but for $^{179}$Hf $(I^{\pi}=k^{\pi}=9/2^+)$.}
\label{fig1_179hf}
\end{figure}
%%%%%%%%%%%%%%%%%%%%%%%%%%%%%%%%%%%%%%%%%%

\clearpage

\newpage

%%%%%%%Fig2-1%%%%%%%%%%%%%%%%%%%%%%%%%%%%%%%%
\begin{figure}[H]
\centering
\includegraphics[width=45mm]{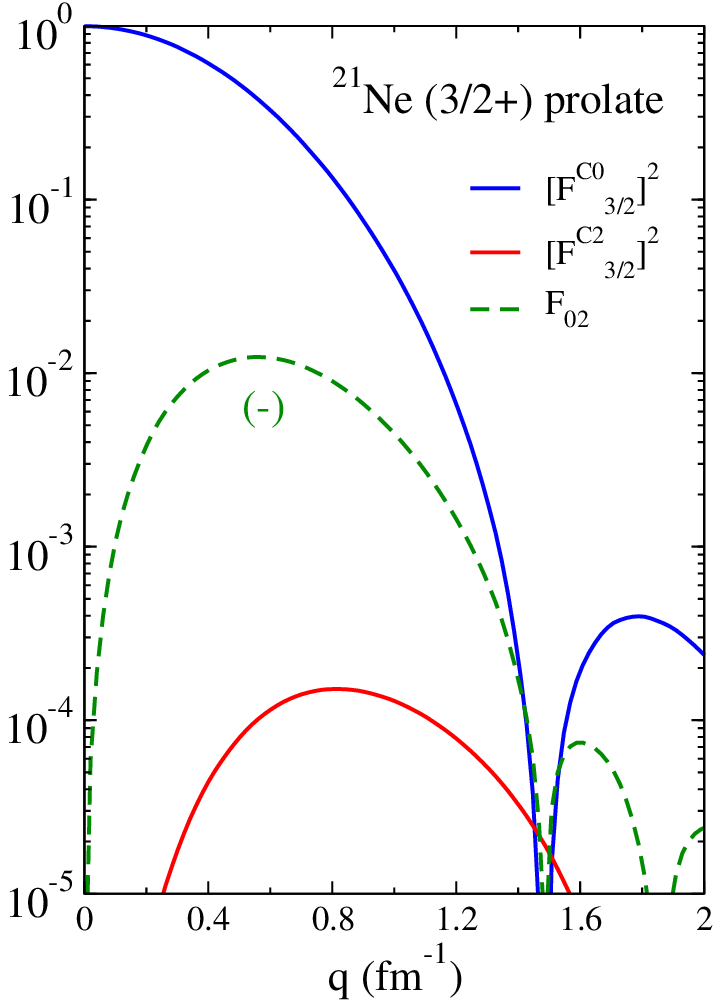}
\caption{Charge form factors $[F^{C0}_{3/2}]^2$ and $[F^{C2}_{3/2}]^2$, as well as $F_{02}$ from Eq. (\ref{f02}) for the prolate shape in $^{21}$Ne.}
\label{fig2_21ne}
\end{figure}
%%%%%%%%%%%%%%%%%%%%%%%%%%%%%%%%%%%%%%%%%%
%%%%%%%Fig2-2%%%%%%%%%%%%%%%%%%%%%%%%%%%%%%%%
\begin{figure}[H]
  \centering
  \includegraphics[width=55mm]{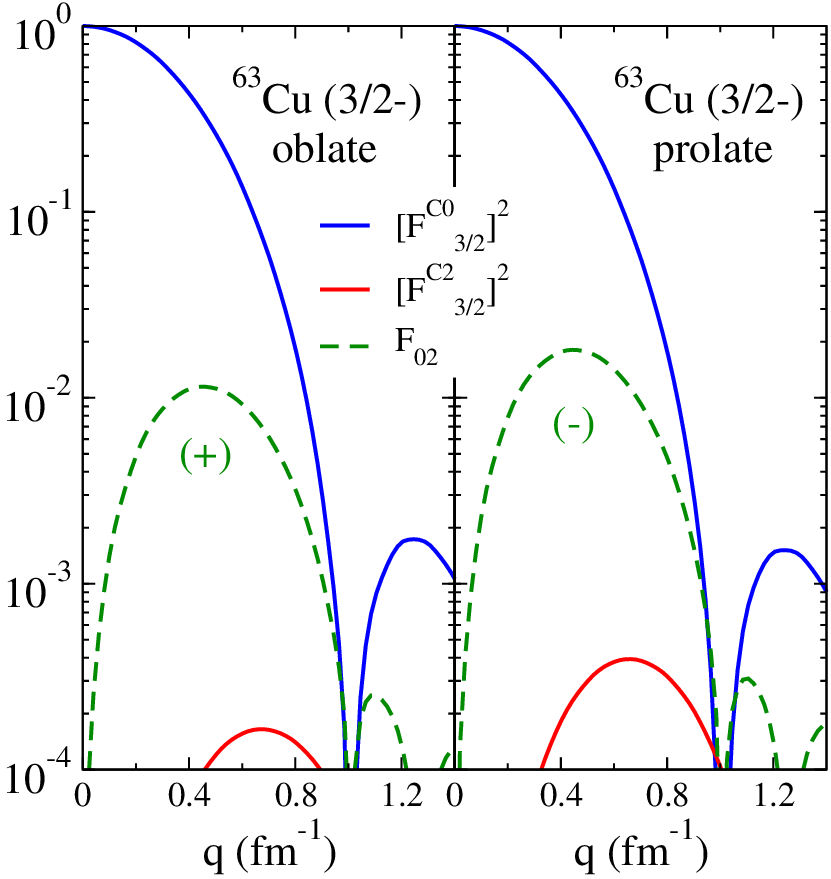}
\caption{Same as in Fig. \ref{fig2_21ne}, but for the oblate and prolate shapes of  $^{63}$Cu.}
\label{fig2_63cu}
\end{figure}
%%%%%%%%%%%%%%%%%%%%%%%%%%%%%%%%%%%%%%%%%%
%%%%%%%%Fig2-3%%%%%%%%%%%%%%%%%%%%%%%%%%%%%%%
\begin{figure}[H]
  \centering
  \includegraphics[width=55mm]{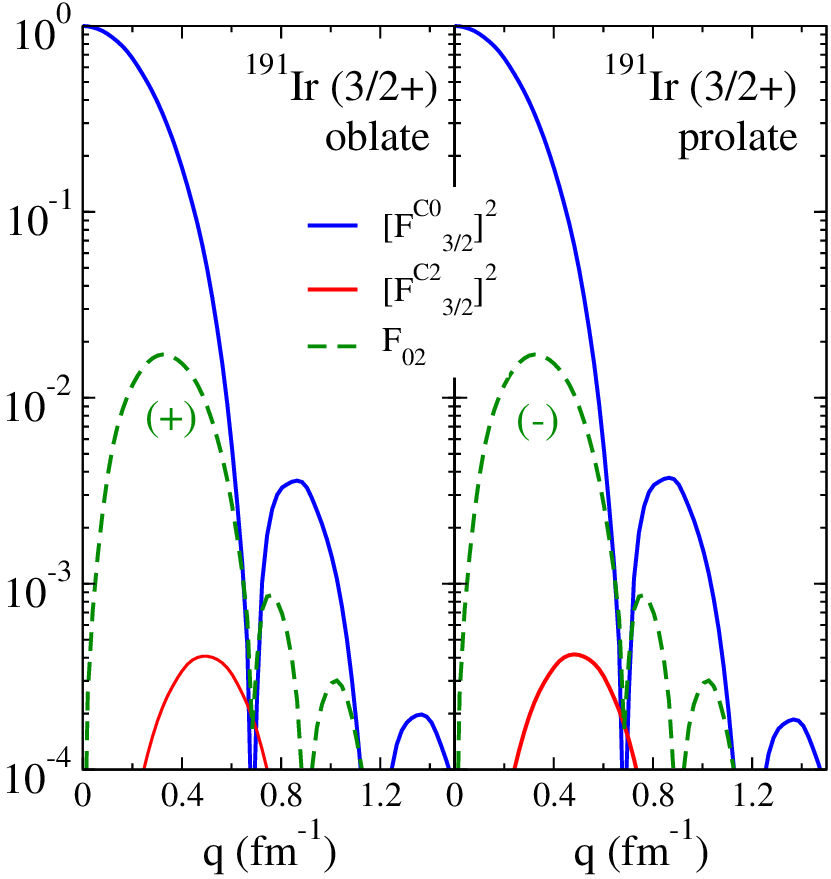}
\caption{Same as in Fig. \ref{fig2_21ne}, but for the oblate and prolate shapes of  $^{191}$Ir.}
\label{fig2_191ir}
\end{figure}
%%%%%%%%%%%%%%%%%%%%%%%%%%%%%%%%%%%%%%%%%%
%%%%%%%Fig2-4%%%%%%%%%%%%%%%%%%%%%%%%%%%%%%%%
\begin{figure}[H]
  \centering
  \includegraphics[width=55mm]{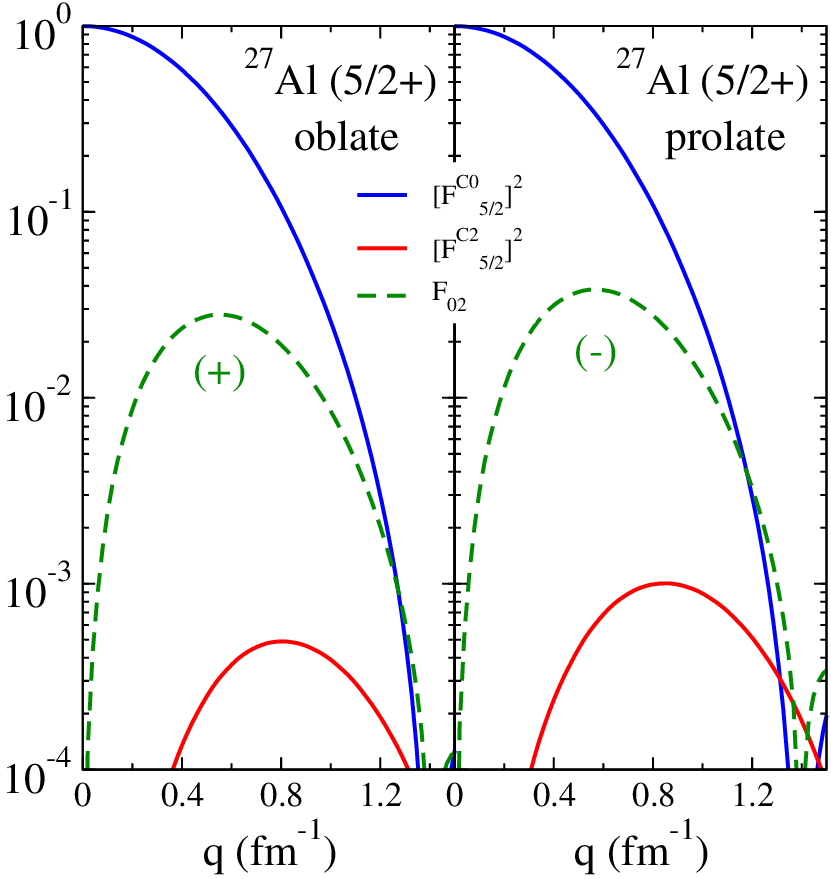}
\caption{Charge $[F^{C\lambda}_{5/2}]^2\, (\lambda=0,2)$ and $F_{02}$ form factors for the oblate and prolate shapes in $^{27}$Al.}
\label{fig2_27al}
\end{figure}
%%%%%%%%%%%%%%%%%%%%%%%%%%%%%%%%%%%%%%%%%%%
%%%%%%%%%Fig2-5%%%%%%%%%%%%%%%%%%%%%%%%%%%%%
\begin{figure}[H]
  \centering
  \includegraphics[width=55mm]{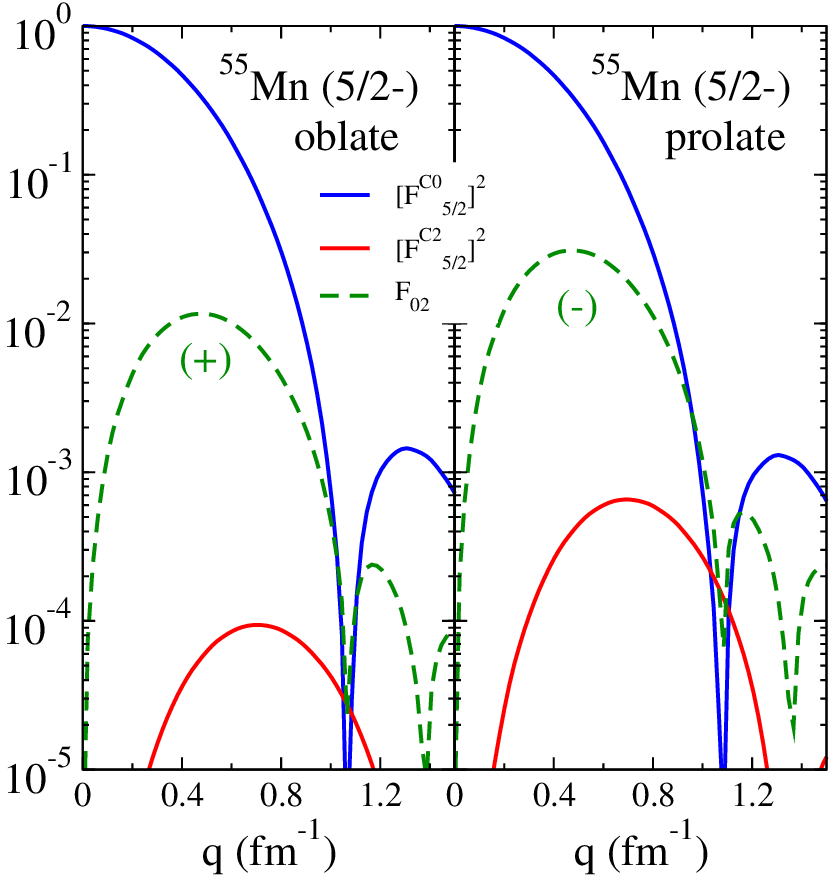}
\caption{Same as in Fig. \ref{fig2_27al}, but for $^{55}$Mn.}
\label{fig2_55mn}
\end{figure}
%%%%%%%%%%%%%%%%%%%%%%%%%%%%%%%%%%%%%%%%%%
%%%%%%%%Fig2-6%%%%%%%%%%%%%%%%%%%%%%%%%%%%%%%%
\begin{figure}[H]
  \centering
  \includegraphics[width=45mm]{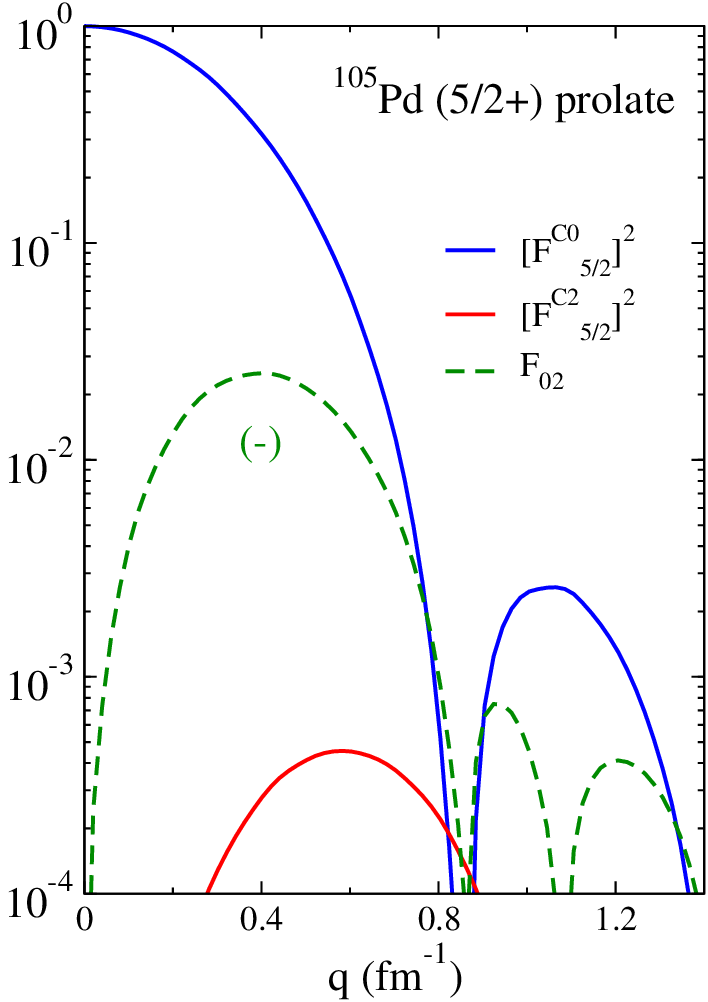}
\caption{Same as in Fig. \ref{fig2_27al}, but for the prolate shape in $^{105}$Pd.}
\label{fig2_105pd}
\end{figure}
%%%%%%%%%%%%%%%%%%%%%%%%%%%%%%%%%%%%%%%%%%%
%%%%%%%%%Fig2-7%%%%%%%%%%%%%%%%%%%%%%%%%%%%%%%
\begin{figure}[H]
  \centering
  \includegraphics[width=45mm]{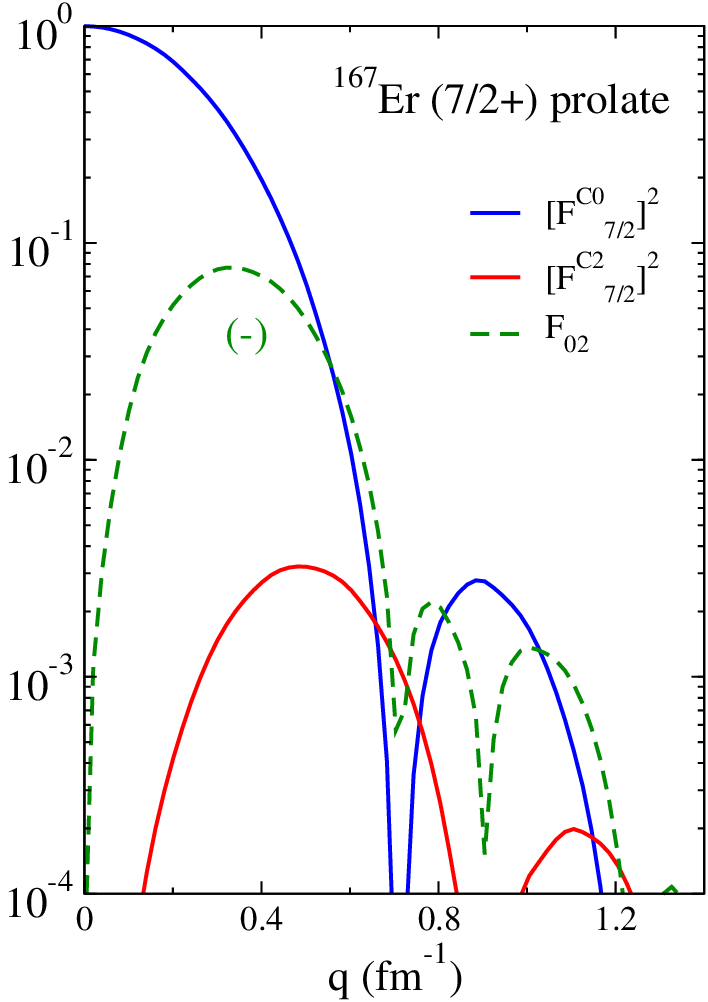}
\caption{Charge $[F^{C\lambda}_{7/2}]^2 \, (\lambda=0,2)$ and $F_{02}$ form factors for the prolate shape in $^{167}$Er.}
\label{fig2_167er}
\end{figure}
%%%%%%%%%%%%%%%%%%%%%%%%%%%%%%%%%%%%%%%%%
%%%%%%%%Fig2-8%%%%%%%%%%%%%%%%%%%%%%%%%%%%%%%%
\begin{figure}[H]
  \centering
  \includegraphics[width=45mm]{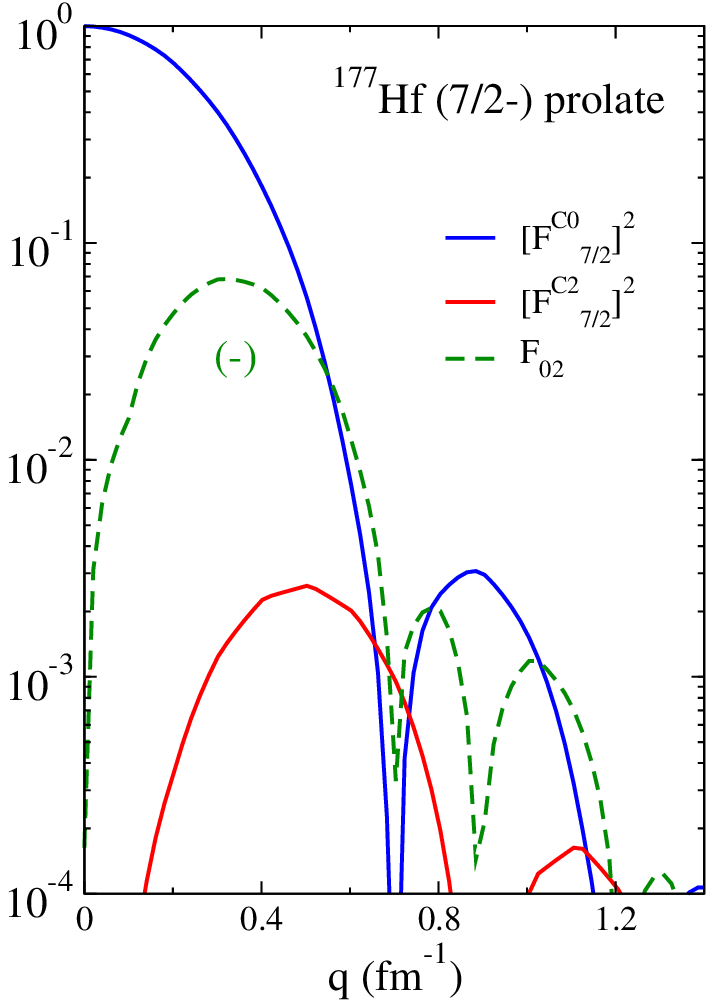}
\caption{Same as in Fig. \ref{fig2_167er}, but for $^{177}$Hf.}
\label{fig2_177hf}
\end{figure}
%%%%%%%%%%%%%%%%%%%%%%%%%%%%%%%%%%%%%%%%
%%%%%%%Fig2-9%%%%%%%%%%%%%%%%%%%%%%%%%%%%%%%
\begin{figure}[H]
  \centering
  \includegraphics[width=45mm]{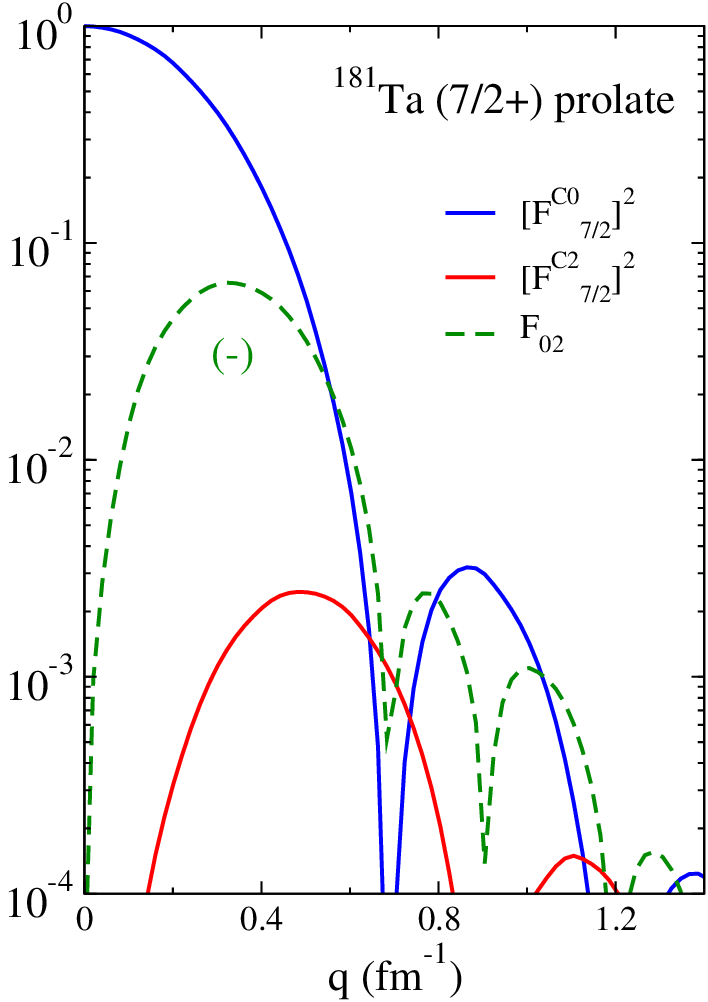}
\caption{Same as in Fig. \ref{fig2_167er}, but for $^{181}$Ta.}
\label{fig2_181ta}
\end{figure}
%%%%%%%%%%%%%%%%%%%%%%%%%%%%%%%%%%%%%%%
%%%%%%%%%Fig2-10%%%%%%%%%%%%%%%%%%%%%%%%%%%%%%%
\begin{figure}[H]
  \centering
  \includegraphics[width=45mm]{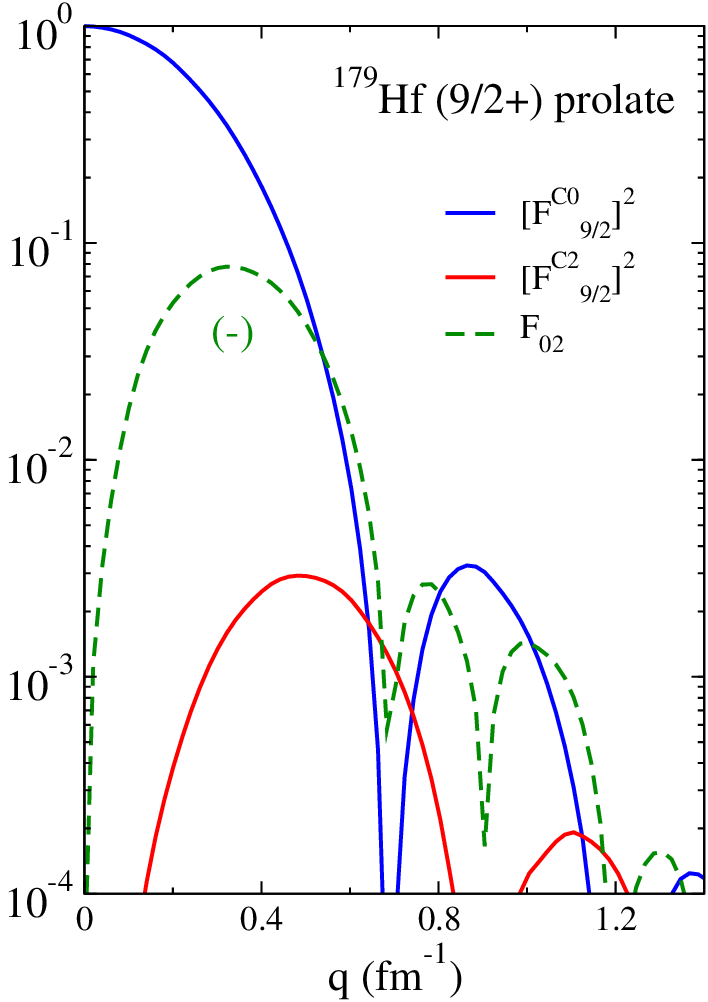}
\caption{Charge $[F^{C\lambda}_{9/2}]^2\, (\lambda=0,2)$ and $F_{02}$ form factors for the prolate shape in $^{179}$Hf.}
\label{fig2_179hf}
\end{figure}
%%%%%%%%%%%%%%%%%%%%%%%%%%%%%%%%%%%%%%%%

\clearpage

\newpage

%%%%Fig3-1%%%%%%%%%%%%%%%%%%%%%%%%%%%%%%%
\begin{figure}[H]
\centering
\includegraphics[width=45mm]{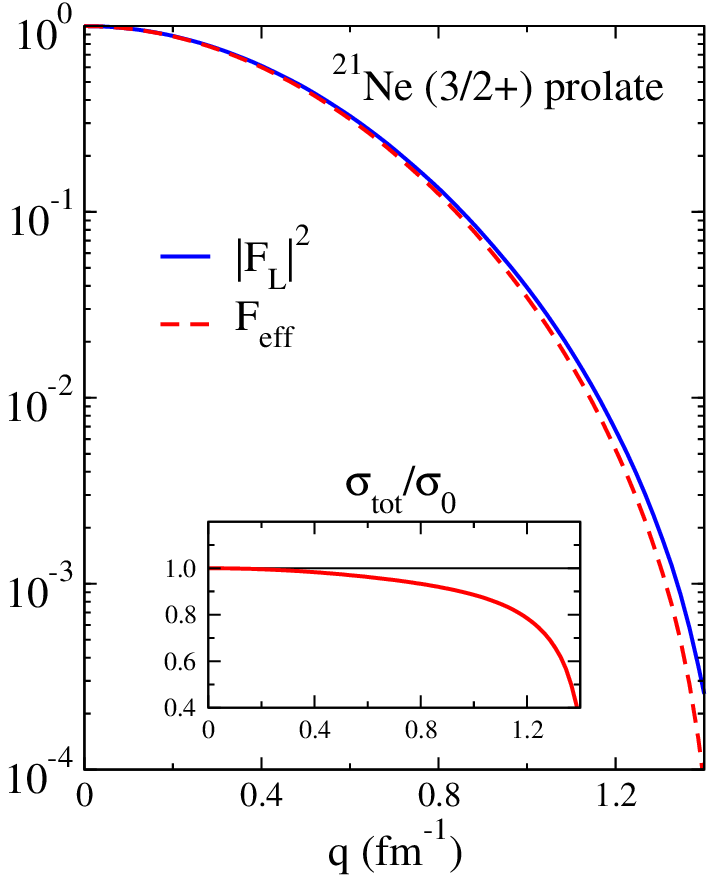}
\caption{$|F_L|^2$ and $F_{\rm eff}$ from Eq. (\ref{feff}) for the prolate shape in $^{21}$Ne. The inset contains the ratio between the total cross-section with aligned nuclei and the unpolarized one.}
\label{fig3_21ne}
\end{figure}
%%%%%%%%%%%%%%%%%%%%%%%%%%%%%%%%%%%%%%%%%
%%%%%%Fig3-2%%%%%%%%%%%%%%%%%%%%%%%%%%%%%%%
\begin{figure}[H]
  \centering
  \includegraphics[width=55mm]{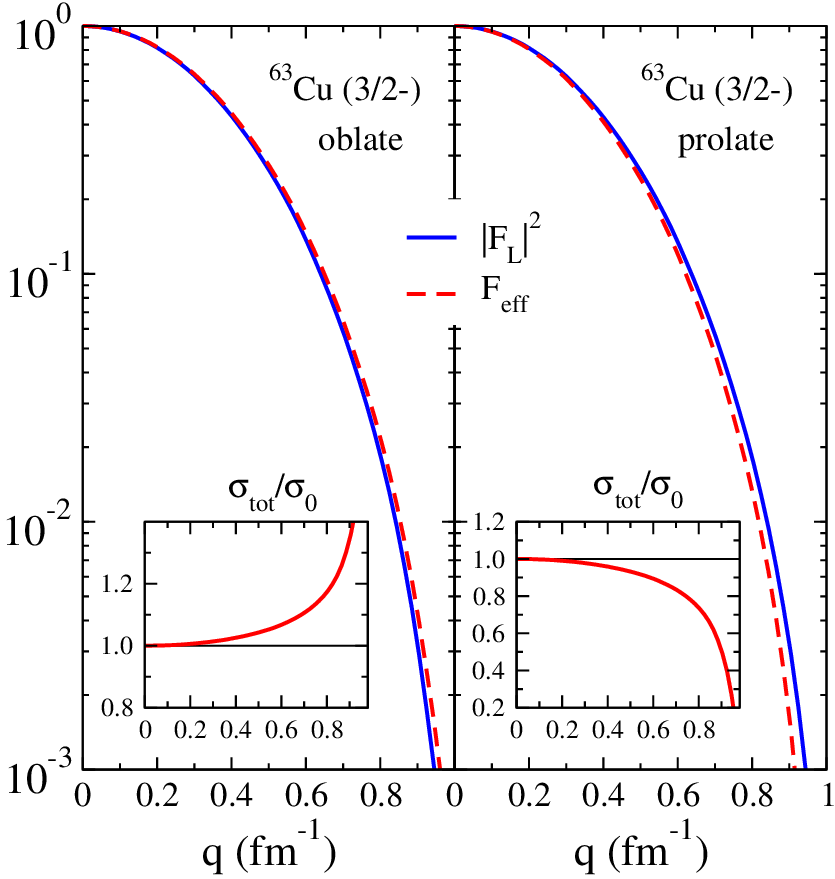}
\caption{Same as in Fig. \ref{fig3_21ne}, but for the oblate and prolate shapes in $^{63}$Cu.}
\label{fig3_63cu}
\end{figure}
%%%%%%%%%%%%%%%%%%%%%%%%%%%%%%%%%%%%%%%%%%
%%%%%%%%Fig3-3%%%%%%%%%%%%%%%%%%%%%%%%%%%%%%%%
\begin{figure}[H]
  \centering
  \includegraphics[width=55mm]{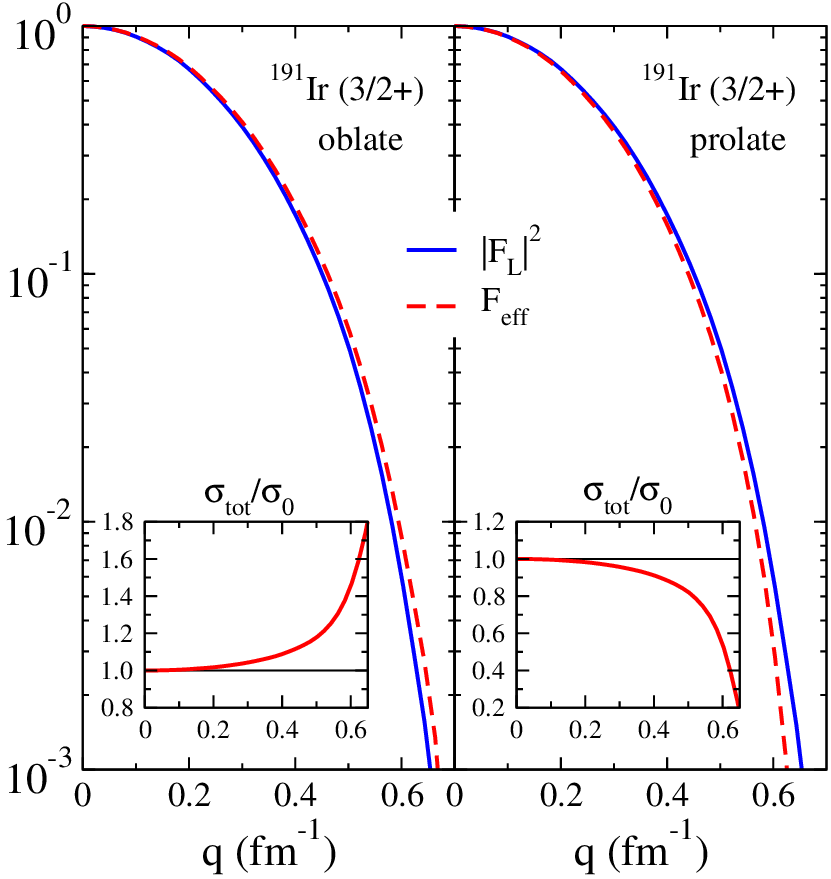}
\caption{Same as in Figure \ref{fig3_21ne}, but for the oblate and prolate shapes in $^{191}$Ir.}
\label{fig3_191ir}
\end{figure}
%%%%%%%%%%%%%%%%%%%%%%%%%%%%%%%%%%%%%%%%%%%
%%%%%%Fig3-4%%%%%%%%%%%%%%%%%%%%%%%%%%%%%%%%
\begin{figure}[H]
  \centering
  \includegraphics[width=55mm]{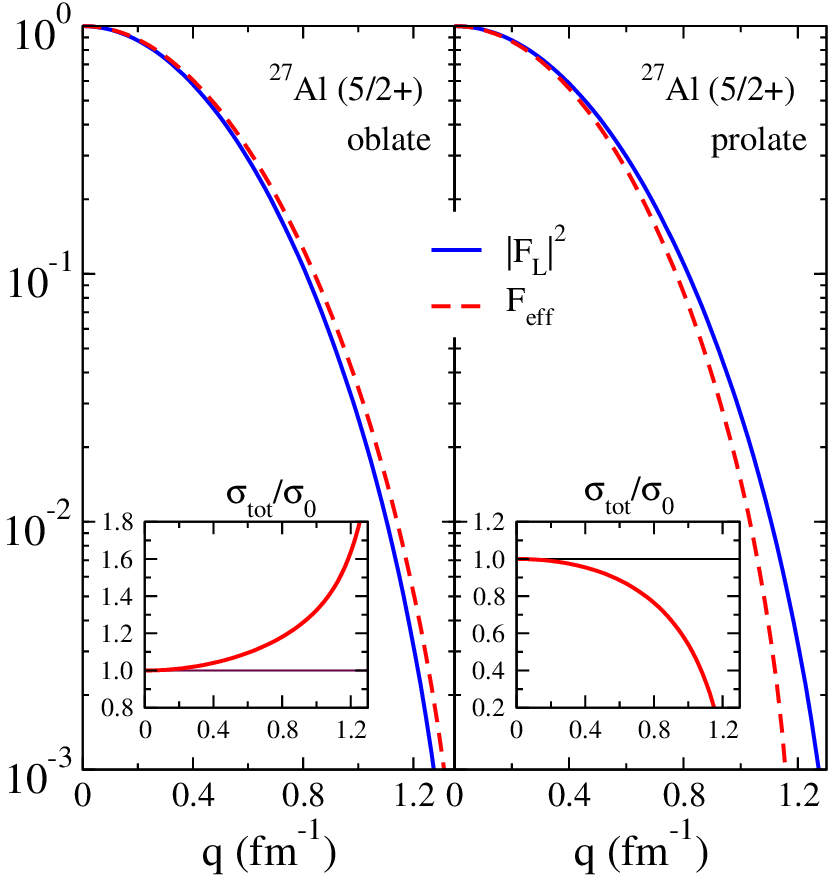}
\caption{Same as in Fig. \ref{fig3_21ne}, but for the oblate and prolate shapes in $^{27}$Al.}
\label{fig3_27al}
\end{figure}
%%%%%%%%%%%%%%%%%%%%%%%%%%%%%%%%%%%%%%%%%%%%
%%%%%%%%%%%Fig3-5%%%%%%%%%%%%%%%%%%%%%%%%%%%%%%%%
\begin{figure}[H]
  \centering
  \includegraphics[width=55mm]{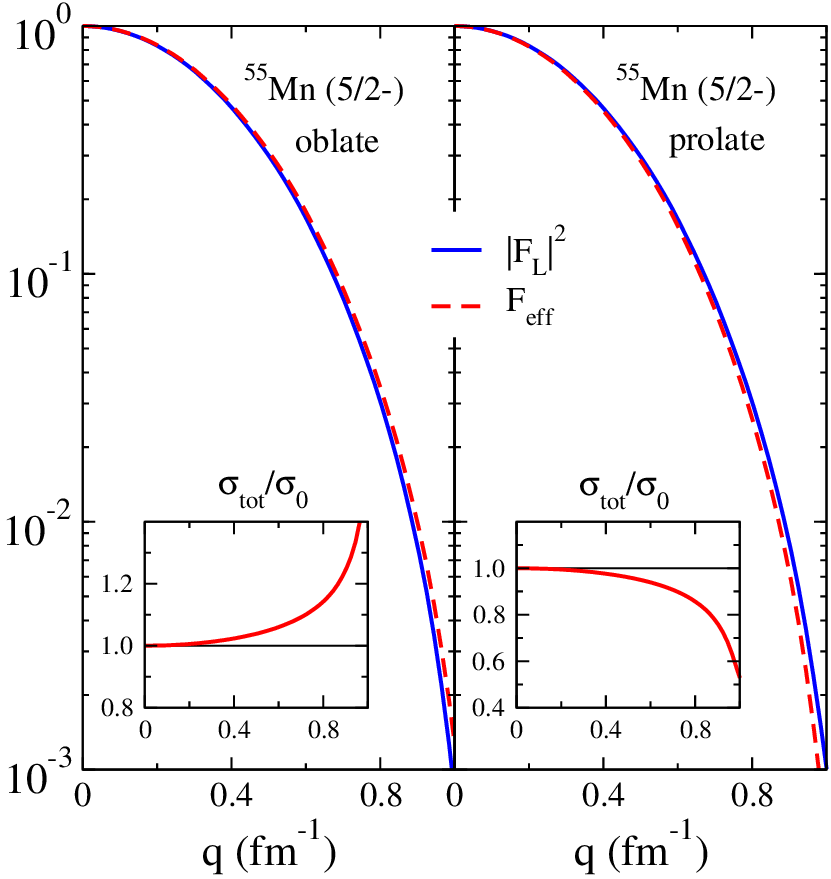}
\caption{Same as in Fig. \ref{fig3_21ne}, but for the oblate and prolate shapes in $^{55}$Mn.}
\label{fig3_55mn}
\end{figure}
%%%%%%%%%%%%%%%%%%%%%%%%%%%%%%%%%%%%%%%%%%%%%
%%%%%%%%Fig3-6%%%%%%%%%%%%%%%%%%%%%%%%%%%%%%%%
\begin{figure}[H]
  \centering
  \includegraphics[width=45mm]{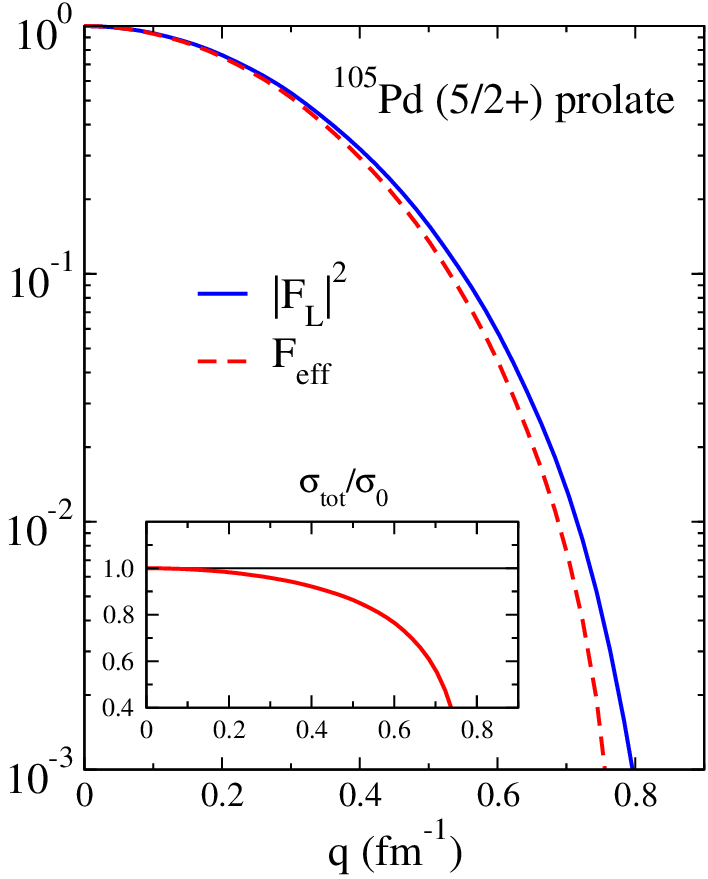}
\caption{Same as in Fig. \ref{fig3_21ne}, but for $^{105}$Pd.}
\label{fig3_105pd}
\end{figure}
%%%%%%%%%%%%%%%%%%%%%%%%%%%%%%%%%%%%%%%
%%%%%%%%%%%Fig3-7%%%%%%%%%%%%%%%%%%%%%%%%%%%%%%%%
\begin{figure}[H]
  \centering
  \includegraphics[width=45mm]{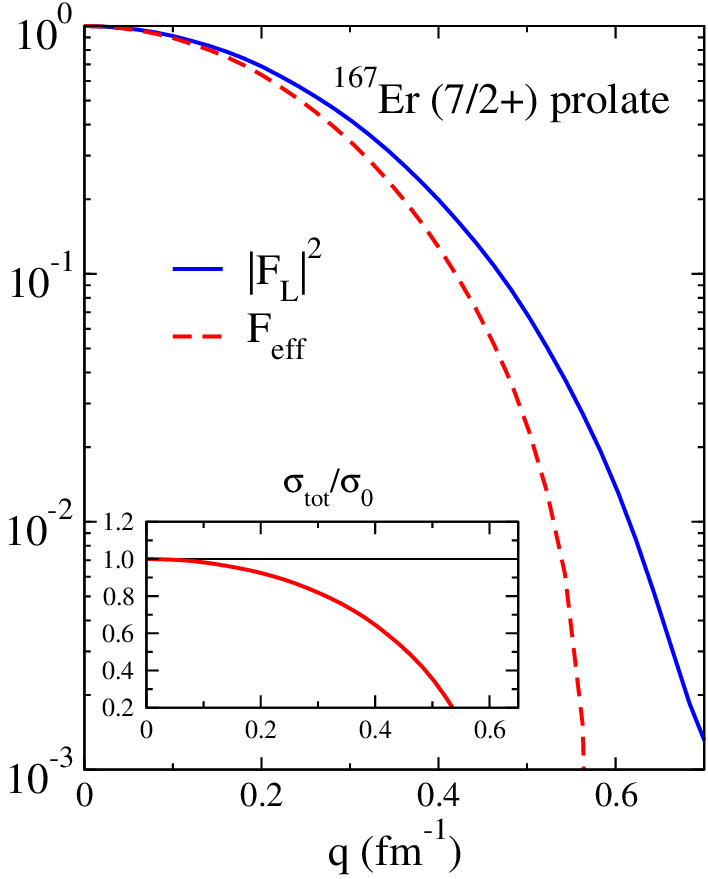}
\caption{Same as in Fig. \ref{fig3_21ne}, but for $^{167}$Er.}
\label{fig3_167er}
\end{figure}
%%%%%%%%%%%%%%%%%%%%%%%%%%%%%%%%%%%%%%%%%%
%%%%%%%%Fig3-8%%%%%%%%%%%%%%%%%%%%%%%%%%%%%%%
\begin{figure}[H]
  \centering
  \includegraphics[width=45mm]{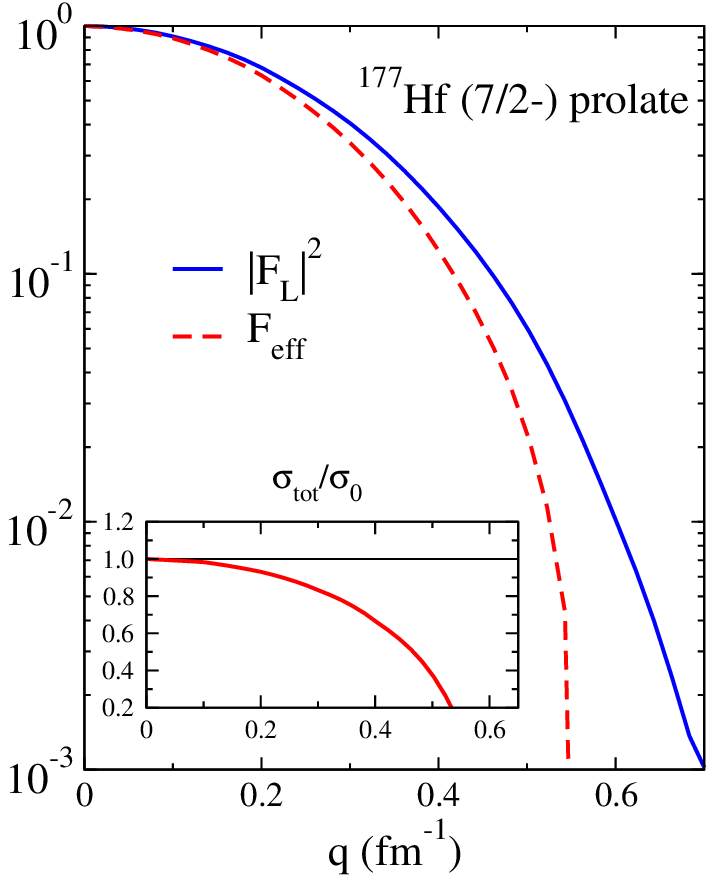}
\caption{Same as in Fig. \ref{fig3_21ne}, but for $^{177}$Hf.}
\label{fig3_177hf}
\end{figure}
%%%%%%%%%%%%%%%%%%%%%%%%%%%%%%%%%%%%%%%%
%%%%%%%%%%%%Fig3-9%%%%%%%%%%%%%%%%%%%%%%%%%%%%%%%
\begin{figure}[H]
  \centering
  \includegraphics[width=45mm]{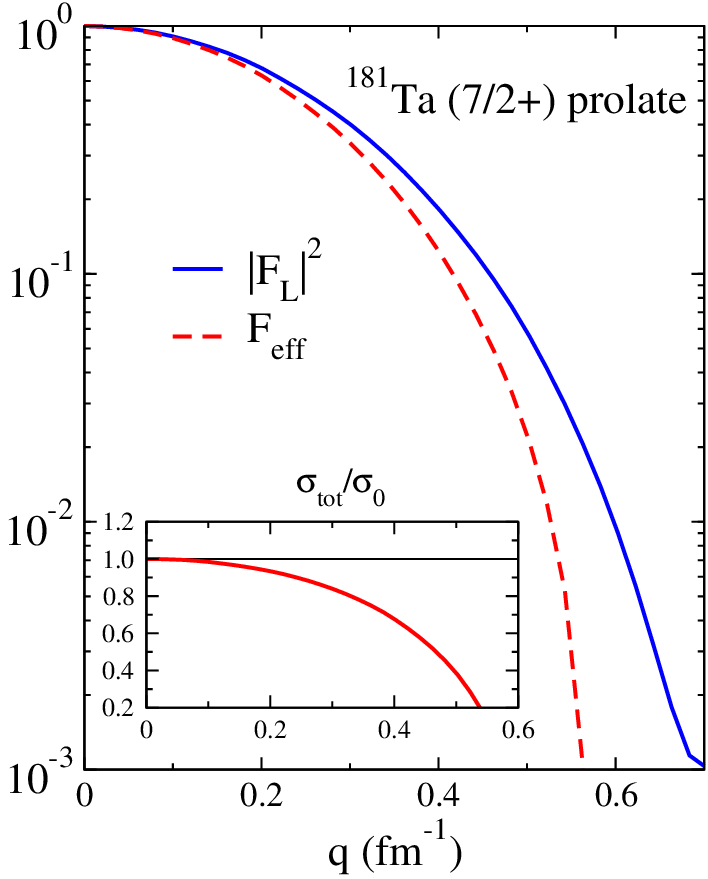}
\caption{Same as in Fig. \ref{fig3_21ne}, but for $^{181}$Ta.}
\label{fig3_181ta}
\end{figure}
%%%%%%%%%%%%%%%%%%%%%%%%%%%%%%%%%%%%%%%%%%%
%%%%%%%%Fig3-10%%%%%%%%%%%%%%%%%%%%%%%%%%%%%%%
\begin{figure}[H]
  \centering
  \includegraphics[width=45mm]{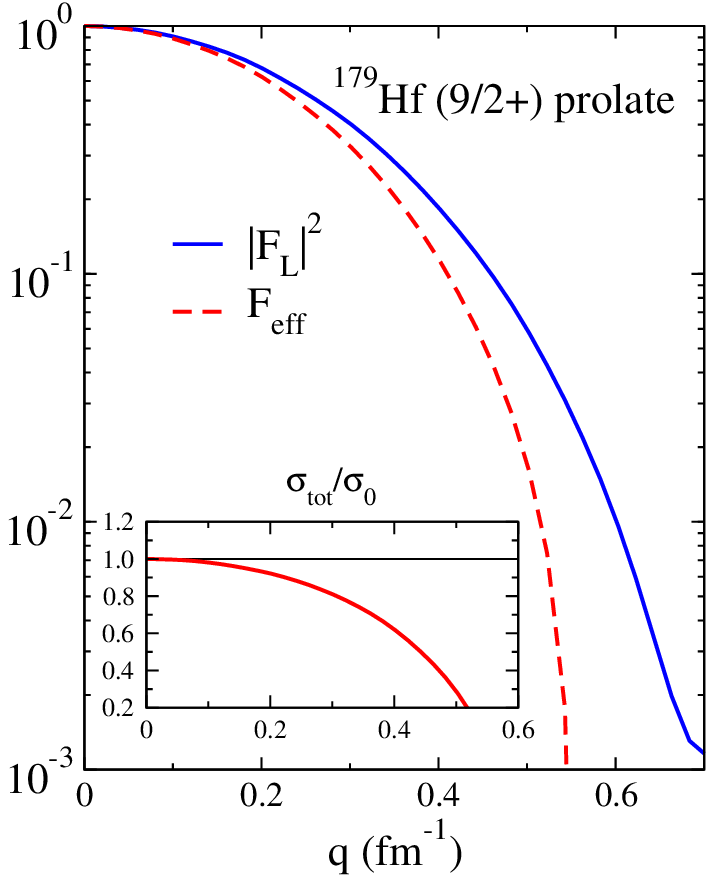}
\caption{Same as in Fig. \ref{fig3_21ne}, but for $^{179}$Hf.}
\label{fig3_179hf}
\end{figure}
%%%%%%%%%%%%%%%%%%%%%%%%%%%%%%%%%%%%%%%%%%

\end{document}